\documentclass[twocolumn]{aastex701}
\usepackage{amsmath,amssymb,bm}
\usepackage{subcaption}

\begin{document}

\title{Effects of Magnetic Fields on Solar Inertial Modes in Global Nonlinear Simulations}

\author[orcid=0009-0004-8465-3881]{Finotti, C.S.} 
\affiliation{Physics Department, Universidade Federal de Minas Gerais, Av. Antonio Carlos, 6627, Belo Horizonte, MG 31270-901, Brazil}
\email{conradofinotti@ufmg.br}

\author[orcid=0000-0002-2671-8796]{Guerrero, G.} 
\affiliation{Physics Department, Universidade Federal de Minas Gerais, Av. Antonio Carlos, 6627, Belo Horizonte, MG 31270-901, Brazil}
\email{guerrero@fisica.ufmg.br}

\author[orcid=0000-0002-7679-3962]{Triana, S. A.} 
\affiliation{Royal Observatory of Belgium, Ringlaan 3, B-1180 Brussels, Belgium}
\email{santiago.triana@observatory.be }

\author[orcid=0000-0002-2227-0488]{Dikpati, M.} 
\affiliation{High Altitude Observatory, NSF  National Center for Atmospheric Research, Boulder, Colorado, USA}
\email{dikpati@ucar.edu}

\author[orcid=0000-0001-7077-3285]{Smolarkiewicz, P.K.} 
\affiliation{ National Center for Atmospheric Research, Boulder, Colorado, USA}
\email{smolar@ucar.edu}

\begin{abstract}
Solar inertial modes are increasingly viewed as probes of the solar interior, with recent helioseismic evidence indicating that these modes may be sensitive to the internal magnetic field. We investigate the role of magnetic fields on inertial mode excitation and dynamics using global nonlinear MHD simulations with differential rotation forced toward a helioseismically derived profile. Initial poloidal fields of varying strengths and  topologies self-consistently generate different toroidal field configurations through the $\Omega$-effect. In all cases, baroclinic instability excites high-latitude inertial modes, with weak fields enhancing its growth rate above the hydrodynamic value. For stronger fields, the response depends on topology and depth.
As a consequence, the polar vortex emerges earlier and modifies its relative amplitude, handedness, and hemispheric asymmetry.
Only magnetic topologies exhibiting toroidal bands at the upper convection zone develop a second, magneto-shear (“edge”) instability at the equatorward edges of these bands. Powered mainly by the toroidal field, this instability sets in earlier and grows faster for stronger initial fields. It excites equatorial modes absent in the hydrodynamic simulation, with fast magneto-Rossby signatures approaching the classical Rossby dispersion relation at tachocline depths. Alfvénic signatures and tentative slow magneto-Rossby branches also appear. 
Inertial-mode excitation and frequencies thus depend on magnetic topology and spatial distribution, not merely field strength.

\end{abstract}

\keywords{Solar interior, Solar inertial modes, Solar rotation, Magnetic field}


\section{Introduction}\label{sec:intro}

Recent advances in helioseismology and surface feature tracking have revealed a rich spectrum of low-frequency global motions in the Sun, prominently including inertial (Rossby-like) modes restored by the Coriolis force \citep{Loptien2018, Hanasoge2019detection, Proxauf2020exploring, Hanson2020solar, Gizon2021, Hanson2022}. These detections highlight several branches, equatorial, critical-layer, high-latitude, and high-frequency, each with distinct dispersion relations, amplitudes, and cycle variability. Complementary observational work identifies correlations and anti-correlations between wave–mean-flow interactions, the emergence of zonal flow patterns, and magnetic cycle variability \citep{WaideleZhao2023, lekshmi2024, Ding2025}. Thus, inertial modes are increasingly recognized as valuable probes of the physical processes operating in the solar interior.

Classically, Rossby modes are described by the dispersion relation $\omega = -2m\Omega/[l(l+1)]$, representing purely toroidal, retrograde motions on a uniformly rotating sphere. In this idealized description, the spherical harmonics are classified as sectoral ($l=m$) or tesseral ($l>m$). The radial vorticity is symmetric (antisymmetric) about the equator for even (odd) $l-m$, while the longitudinal velocity has the opposite parity. In the Sun, the observed equatorial Rossby branch approximately follows the sectoral form of this relation.

At higher latitudes, Rossby-like branches deviate from the classical spectrum: high-latitude modes appear as spiral structures near the poles, dominated by the \(m=1\) component \citep{hathaway2021hydrodynamic, Gizon2021, Ding2025, LiangGizon2025}, and their frequencies scale approximately as \(\omega \simeq m\Omega_2\) under differential rotation, where $\Omega_2$ approximately characterizes the equator-to-pole rotational shear \citep{fournier2022viscous}. Theoretical and numerical results suggest that these modes are sensitive to latitudinal gradients of potential temperature \citep{Bekki2022a, Bekki2024,SouzaGomes2026}. Other families of inertial modes include critical-layer modes (often termed ``critical-latitude'' modes)—peaking where the pattern speed matches the local rotation rate—and high-frequency modes that are antisymmetric about the equator and oscillate roughly three times faster than sectoral equatorial modes \citep{Hanson2022, Triana2022}.

Magnetohydrodynamic extensions of the shallow-water framework (MHD-SWT) indicate that at tachocline levels, the Rossby waves become modified by the Lorentz force into magneto-Rossby waves \citep{Zaqarashvili2007, Zaqarashvili2009, Zaqarashvili2018}. Nonlinear MHD-SWT simulations further reveal quasi-periodic tachocline oscillations sustained by cyclic exchanges among kinetic, magnetic, and potential energy reservoirs through a network of conversion processes involving Reynolds, Maxwell, and mixed stresses \citep{Dikpati2018}. More recently, incorporating penetrative convection into the MHD-SWT equations through an $\alpha$-effect, has been shown, through linear wave analysis and numerical solutions of the resulting dispersion relation, to produce coupled Rossby–dynamo modes. These modes allow an exchange between rotational and convective energy reservoirs and yield characteristic timescales consistent with both Schwabe and Rieger-type periodicities \citep{Zaqarashvili2025}.

On the observational side, \citet{HanasogeHanson2026} recently reported the possible helioseismic detection of magnetically modified Rossby modes in the Sun. Their two-dimensional analysis cannot localize the field in radius: the inferred amplitude of $\sim5\times10^{3}$~G corresponds to the base of the convection zone only under the assumption that the field is anchored there, while the observations primarily probe the outer convection zone. Disentangling the depth at which the field shapes these modes is among the questions that motivate the fully three-dimensional MHD simulations developed here.

Our previous hydrodynamic simulations \citep[hereafter Paper I]{SouzaGomes2026} established a baseline for the excitation and nonlinear evolution of solar inertial modes in the absence of magnetic fields. Simulations initialized with individual spherical harmonic perturbations reproduced the classical Rossby dispersion relation. The experiments also revealed nonlinear mode coupling through direct and inverse energy cascades and identified a baroclinic instability as the main excitation mechanism of high-latitude retrograde vortices. In regimes of strong shear, the resulting Reynolds stresses were able to feed back on the mean flow, accelerating the rotation of the polar regions and modulating differential rotation. 

Despite this progress, a fully three-dimensional assessment of how large-scale magnetic fields reshape inertial modes in a solar-like shell remains incomplete. Building upon the hydrodynamic results of Paper I, the inclusion of magnetic fields raises several questions that must be addressed before the observed mode properties can be used to infer the conditions of the solar interior. How does magnetic tension interact with rotational shear and baroclinicity in controlling the stability and characteristic frequencies of inertial modes? How does the magnetic-field topology modify the spatial structure and growth of the high-latitude modes? Can a large-scale magnetic field promote equatorial retrograde branches that are absent or weak in the hydrodynamic regime?

In this work, we address these questions using nonlinear global MHD simulations of a solar-like interior. Relative to Paper I, the target differential-rotation profile is now constructed from helioseismic inversions, providing a more realistic radial and latitudinal distribution of rotational shear. Because the large-scale magnetic configuration inside the Sun remains poorly constrained, we consider initial poloidal fields with different strengths and topologies and examine how these properties affect the subsequent dynamics.

In most of the current literature, magneto-Rossby modes are commonly studied using prescribed toroidal magnetic fields. In the simulations presented here, instead, the toroidal component develops self-consistently from the initial poloidal configuration through the $\Omega$-effect driven by differential rotation. Its spatial distribution is therefore not imposed in advance, but emerges from the interaction between the differential rotation and the magnetic topology. This approach allows a direct comparison between the HD and MHD regimes and enables us to investigate how magnetic fields affect the excitation, growth, frequency spectra, and morphology of solar-like inertial modes.

The paper is organized as follows. In Section~\ref{sec:methods}, we describe the numerical model, the calibrated differential-rotation profile, and the initial magnetic-field configurations. In Section~\ref{sec:results}, we first present the hydrodynamic reference model and then examine how the magnetic-field strength and topology influence the mean state, the evolution of the kinetic and magnetic energies, the growth of non-axisymmetric instabilities, the spectrum of excited inertial modes, and the morphology of the polar vortex. Section~\ref{sec:conclusions} summarizes the main conclusions.

\section{Numerical Model}
\label{sec:methods}

The numerical framework employed in this work builds directly upon the hydrodynamic setup of Paper I, extended to include magnetic fields using the EULAG-MHD code \citep{Ghizaru2010, SmolarkiewiczCharbonneau2013}. The code solves the magnetohydrodynamic extension of the Lipps--Hemler anelastic system of equations \citep{LippsHemler1982} in a rotating spherical shell. The computational domain spans $0.6\,R_\odot \le r \le 0.96\,R_\odot$ in radius, $0 \le \theta \le \pi$ in colatitude, and $0 \le \phi \le 2\pi$ in longitude. 

The governing equations retain the anelastic mass-conservation and potential-temperature equations employed in Paper I, while the momentum equation is extended to include the Lorentz force and is coupled to the magnetic induction equation. For completeness, the full system solved in the present simulations is written as

\begin{equation}
\nabla\cdot\left(\rho_{\rm ad}\mathbf{u}\right)=0,
\label{eq:continuity}
\end{equation}

\begin{equation}
\frac{D\mathbf{u}}{Dt} + 2\mathbf{\Omega} \times \mathbf{u} = 
- \nabla \left( \frac{p'}{\rho_{\rm ad}} \right) 
+ \mathbf{g}\,\frac{\Theta'}{\Theta_{\rm ad}} 
+ \frac{1}{\mu_0 \rho_{\rm ad}} (\mathbf{B}\cdot\nabla)\mathbf{B} 
- \mathbf{F}_\phi,
\label{eq:momentum}
\end{equation}

\begin{equation}
\frac{D\Theta'}{Dt}
=
-\mathbf{u}\cdot\nabla\Theta_{\rm amb}
-\frac{\Theta'}{\tau},
\label{eq:potential_temperature}
\end{equation}

\begin{equation}
\frac{D\mathbf{B}}{Dt} = (\mathbf{B}\cdot\nabla)\mathbf{u} 
- \mathbf{B}(\nabla\cdot\mathbf{u}).
\label{eq:induction}
\end{equation}

Here, $D/Dt=\partial/\partial t+\mathbf{u}\cdot\nabla$ is the material derivative, $\mathbf{u}$ is the velocity in the rotating frame, and $\mathbf{B}$ is the magnetic field. The vectors $\boldsymbol{\Omega}$ and $\mathbf{g}$ denote the angular velocity of the reference frame and the gravitational acceleration, respectively. The quantities $\rho_{\rm ad}$ and $\Theta_{\rm ad}$ are the density and potential temperature of the hydrostatic adiabatic reference state, while $\Theta'=\Theta-\Theta_{\rm amb}$ denotes departures from the prescribed ambient potential temperature. The potential-temperature perturbation $\Theta'$ is evolved dynamically and relaxed toward the ambient state on the timescale $\tau$. The pressure perturbation $p'$ now includes the magnetic pressure $B^2/(2\mu_0)$, where $\mu_0$ is the vacuum permeability. The term $\mathbf{F}_{\phi}$ represents the longitudinal relaxation forcing defined below.

The adiabatic reference and ambient thermodynamic states are constructed from the polytropic relations introduced in Paper I. The ambient polytropic index varies smoothly with radius following an error function (their Eq.~6), so that the domain comprises a strongly sub-adiabatic radiative interior, a transition layer, and a mildly sub-adiabatic upper region with a polytropic index of $1.5001$. Two refinements with respect to the hydrodynamic baseline are introduced here in order to better match recent helioseismic constraints. First, the radius of the adiabaticity transition is shifted slightly outward, from $0.70\,R_\odot$ in Paper I to $0.71\,R_\odot$ in the present setup. This places the steep variation of $|\delta| = |\nabla_{\rm amb} - \nabla_{\rm ad}|$ slightly above the center of the tachocline, so that the strong radial shear is fully embedded in a stiff, sub-adiabatic layer, consistent with the current understanding of the solar interior. Here, $\nabla = (p/T)\,(dT/dp)$ is the logarithmic temperature gradient with respect to pressure, and the subscripts $\textbf{amb}$ and $\textbf{ad}$ refer to the ambient and adiabatic reference states, respectively. 

\begin{figure}[h]
\centering
\includegraphics[width=1.0\linewidth]{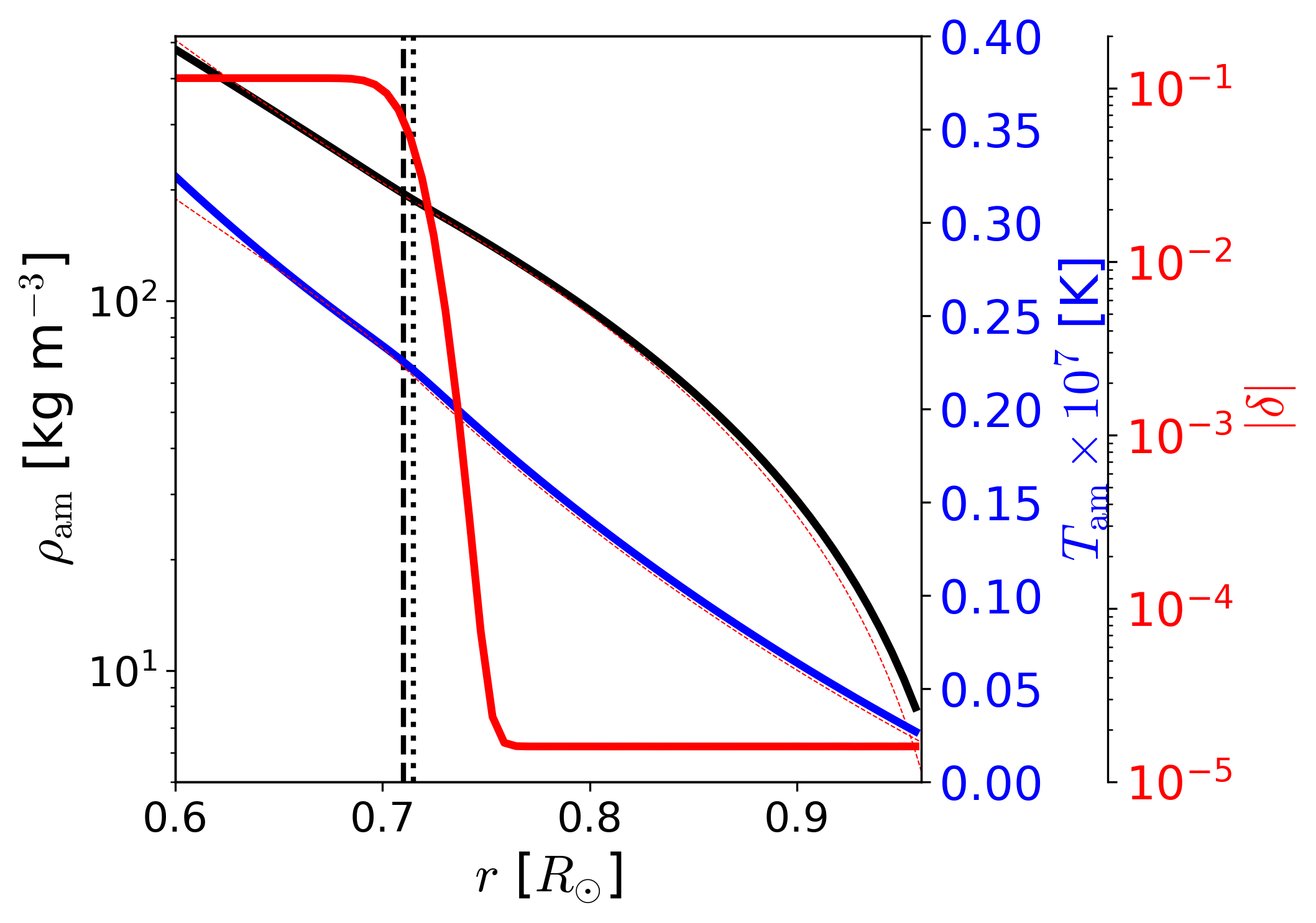}
\caption{Thermodynamic structure of the simulations. The black and blue 
solid lines show, respectively, the density and temperature of the ambient state as a function of radius. The thin (faint) red dashed lines correspond to the same quantities in the solar model of \citep{ChristensenDalsgaard1996}. The thick red line displays the absolute value of the adiabaticity, $|\delta| = |\nabla_{\rm amb} - \nabla_{\rm ad}|$. The vertical dashed and dotted lines mark, respectively, the transition between the highly and slightly sub-adiabatic layers (at $r = 0.71\,R_\odot$) and the location of the tachocline derived from helioseismic inversions (at $r \simeq 0.715 \,R_\odot$).}
\label{fig:strat}
\end{figure}

Figure~\ref{fig:strat} shows the density (black), temperature  (blue), and adiabaticity (red) profiles corresponding to the ambient state. The thin red dashed lines correspond to the solar model of \citet{ChristensenDalsgaard1996}, demonstrating that our stratification is in good agreement with the standard solar model. The vertical dashed and dotted lines mark the transition between the highly and slightly sub-adiabatic layers and the location of the tachocline, respectively.

The second refinement regards the target differential rotation profile, $u_F(r,\theta) = r\sin\theta\,\Omega_F(r,\theta)$, that enters the longitudinal forcing term,
\begin{equation}
\mathbf{F}_\phi = \frac{u_\phi - u_F}{\tau_s}\,\hat{\mathbf{e}}_\phi,
\label{eq:forcing}
\end{equation}
Unlike in Paper I, we now construct this term using a rotation-rate profile $\Omega_F(r,\theta)$ derived from helioseismic inversions of observations from the Solar and Heliospheric Observatory (SOHO) and the Solar Dynamics Observatory (SDO), averaged over the period 1996–2024 \citep{Rozelot2024}. The resulting profile is then mirrored across the equator.
This procedure delivers a more realistic mean flow, including a tachocline whose thickness and radial position are observationally constrained, as well as the latitudinal dependence of the rotation rate within the convection zone. The relaxation time scale is kept at $\tau_s = 1.3\times10^7$~s, corresponding to roughly five Carrington rotations, ensuring consistency with the hydrodynamic baseline while allowing non-axisymmetric perturbations to develop along the imposed mean flow.

Simulations are performed for both purely hydrodynamic and magnetized cases. In the latter, a large-scale poloidal magnetic field is imposed at initialization and the toroidal component is left to develop self-consistently through the $\Omega$-effect sustained by the differential rotation. 

Similarly to \citet{SuleRudiger2005}, the initial poloidal field can be represented by the flux function $A(r,\theta)$,
\begin{equation}
\mathbf{B} = 
\left(
\frac{1}{r^2\sin\theta}\frac{\partial A}{\partial\theta},\;
-\frac{1}{r\sin\theta}\frac{\partial A}{\partial r},\;
0
\right),
\label{eq:poloidal_field}
\end{equation}
with
\begin{equation}
A(r,\theta) = - S_0\,r^2
\left(1-\frac{r}{r_{\rm out}}\right)
\left(1-\frac{r_{\rm in}}{r}\right)
\sin^2\!\theta,
\label{eq:poloidal_potential}
\end{equation}
where $S_0$ controls the field amplitude and $r_{\rm in}$, $r_{\rm out}$ are the inner and outer radii at which the radial component of the field vanishes. By construction, this formulation guarantees $\nabla\cdot\mathbf{B}=0$. We define the initial field amplitude parameter $B_0$ through $B_0=S_0/10$ (in SI units), calibrated for the closed-field configuration, and apply the same conversion to all topologies, so that $B_0$ should be regarded as a common amplitude label rather than as an identical local field strength across the three topologies.

\begin{figure}
\centering
\includegraphics[width=1.0\linewidth]{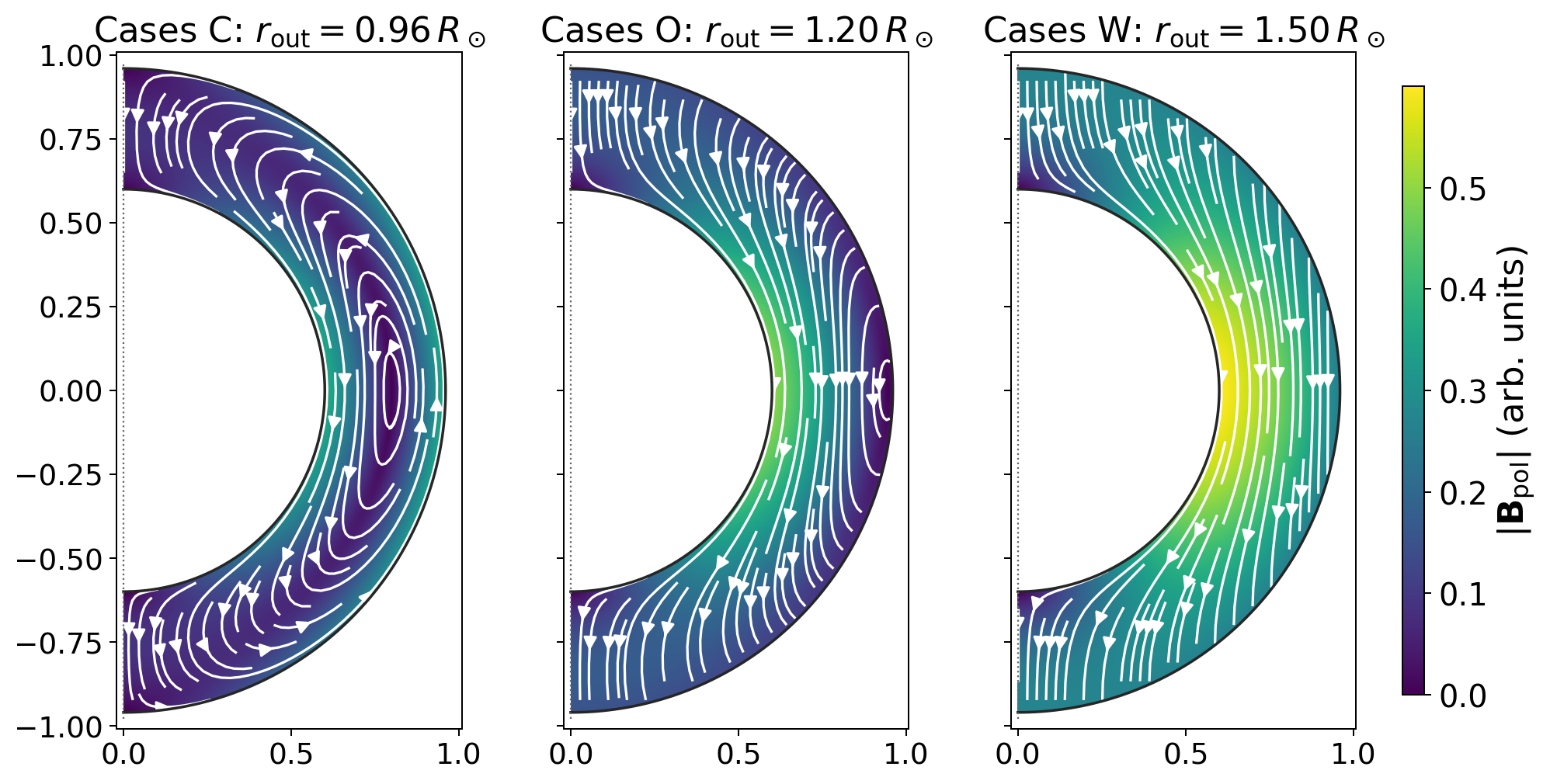}

\caption{Initial poloidal magnetic field configurations in the meridional plane for the three families of magnetic initial conditions. White streamlines trace $\mathbf{B}_{\rm pol}$ ($\mathbf{B}_{\rm pol}=B_r\hat{\mathbf e}_r+B_\theta\hat{\mathbf e}_\theta$) and the color map shows its magnitude in arbitrary units. From left to right: Cases C ($r_{\rm out} = 0.96\,R_\odot$), closed field lines; Cases O ($r_{\rm out} = 1.20\,R_\odot$), open field lines; and Cases W ($r_{\rm out} = 1.50\,R_\odot$), wide-open topology.}
\label{fig:topologies}
\end{figure}

By changing the parameter $r_{\rm out}$ we explore three distinct field topologies, illustrated in Fig.~\ref{fig:topologies}. In Cases C (closed), $r_{\rm out} = 0.96\,R_\odot$, forming a dipolar configuration mostly confined to the simulated volume. In Cases O (open), $r_{\rm out} = 1.20\,R_\odot$, the poloidal field lines cross the radial domain mimicking a moderately open magnetosphere. Finally, in Cases W (wide-open), $r_{\rm out} = 1.50\,R_\odot$, yielding a field with roughly vertical field lines inside the simulation domain. 

For the velocity field we adopt impermeable, stress-free conditions at both radial boundaries. Thermal boundary conditions impose vanishing enthalpy flux at the top and bottom of the domain. The magnetic field is assumed to be radial at both radial boundaries. The computational domain is discretized on a \(128\times64\times64\) grid in longitude, colatitude, and radius, respectively. The thermal-relaxation timescale in Eq.~(\ref{eq:potential_temperature}) is set equal to the longitudinal forcing timescale, $\tau=\tau_s=1.3\times10^7\,{\rm s}$. Apart from the modifications described above, the remaining model parameters and numerical settings follow Paper I, which provides further details of the thermodynamic formulation and numerical implementation.

\section{Results}\label{sec:results}

In the following subsections we present the results of our nonlinear MHD simulations, organized in a way to better comprehend the role of the magnetic field in the dynamics of inertial modes. We first describe the calibrated hydrodynamic simulation that serves as the baseline for the MHD simulations (Section~\ref{sec:HD_model}). We then analyze the MHD cases (Section~\ref{sec:MHD_results}), examining the influence of the field topology, the energy evolution, the spectra of excited inertial modes, and the impact of the magnetic field on the solar polar vortex characterized in Paper I. In Section~\ref{sec:conclusions}, we summarize our findings. 

\subsection{HD calibrated model}\label{sec:HD_model}

\begin{figure}                 
\begin{center}
    \includegraphics[width=1.0\columnwidth]{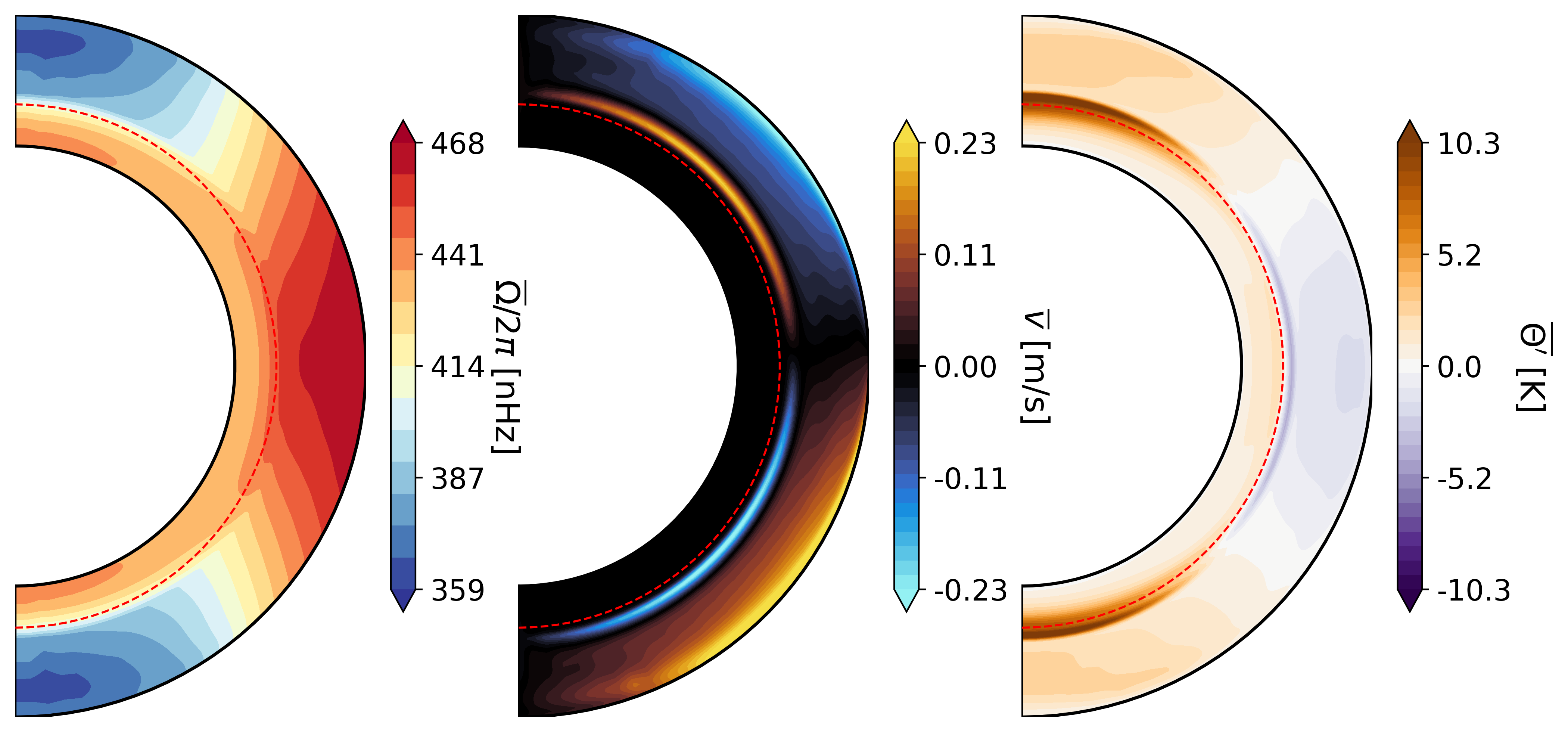}
    \caption{Meridional profiles of the angular velocity, $\overline{\Omega}/2\pi$ (left column), the latitudinal velocity, $\overline{v}$ (middle), and perturbations of potential temperature, $\overline{\Theta^{\prime}}$ (right), for the HD baseline simulation.
    The thin red dashed line marks the center of the tachocline.}
    \label{fig:meridional_prof}
\end{center}
\end{figure}

We first run a purely hydrodynamic simulation of the calibrated model introduced in Section~\ref{sec:methods}, which serves as the baseline against which the MHD cases will be compared. Figure~\ref{fig:meridional_prof} shows, from left to right, the time and azimuthally averaged differential rotation $\overline{\Omega}/2\pi$, latitudinal velocity, $\overline{v}$, representative of the meridional circulation, and the potential temperature perturbation, $\overline{\Theta'}$, in the meridional plane. The temporal average is taken over a later phase of the simulation, when the mean flows evolve more slowly. The red dashed line marks the location of the tachocline. This figure is the analog of Fig.~9 of Paper I.

The relaxed differential rotation exhibits a pole-to-equator shear of $\sim 23\%$, slightly above that of the solar-like ($20\%$ shear) case of Paper I. The rotation contours are tilted relative to the rotation axis, showing clear departures from cylindrical rotation within the convection zone.

The imposed differential-rotation forcing drives a meridional circulation with one dominant cell in each hemisphere. Its equatorward branch is concentrated in a thin layer above the tachocline, while the poleward return flow extends through the upper convection zone. The potential temperature perturbation exhibits its strongest latitudinal gradients near the tachocline, with much weaker baroclinicity throughout the bulk of the convection zone.

The concentration of baroclinicity near the tachocline is consistent with the thermal wind balance associated with the imposed rotational shear (Thermal wind balance, Eq.~12 of Paper I). This, in turn, triggers the baroclinic instability whose nonlinear evolution excites high-latitude inertial modes, whose manifestations include a polar vortex. The properties of these modes will be presented in the following sections compared to the results of the MHD simulations.

\subsection{MHD Results}\label{sec:MHD_results}

The MHD simulations comprise the three poloidal field topologies described in Section~\ref{sec:methods} (Cases C, O, and W), for initial field amplitudes spanning $B_0 = [5,10,15,20]$ Gauss. Each case is compared against the calibrated hydrodynamic baseline of Section~\ref{sec:HD_model}, allowing us to isolate the imprint of the magnetic field on the structure, energetics, and spectrum of the inertial modes. We first examine how the field topology shapes the mean state and the self-consistently generated toroidal field (Section~\ref{sec:field_topology}), then quantify the energy evolution (Section~\ref{sec:energy_evolution}), characterize the spectra of excited modes (Section~\ref{sec:inertial_modes}), and finally assess the impact of the field on the solar polar vortex (Section~\ref{sec:polar_vortex}).

\subsubsection{Influence of Field Topology}\label{sec:field_topology}

\begin{figure}                 
\begin{center}
    \includegraphics[width=1.0\columnwidth]{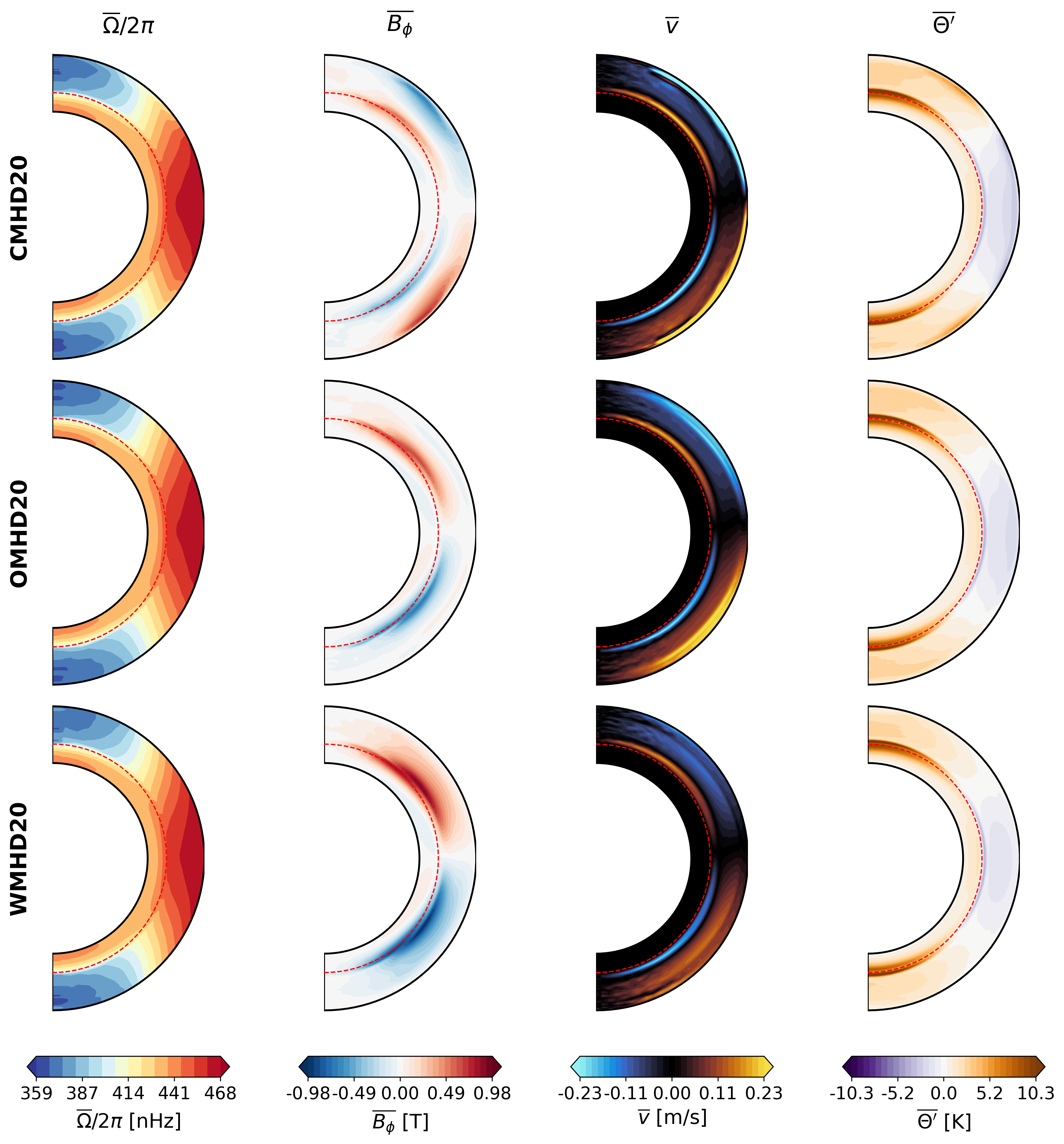}
    \caption{Time and azimuthally averaged meridional view of the differential rotation $\overline{\Omega}/2\pi$, the toroidal magnetic field $\overline{B}_{\phi}$, the latitudinal velocity, $\overline{v}$,  and the potential-temperature perturbation $\overline{\Theta}'$, from left to right for an MHD simulation with $B_0=20$ G. Rows correspond, from top to bottom, to simulations CMHD20 (closed), OMHD20 (open), and WMHD20 (wide-open).  The red dashed curve marks the center of the tachocline.}

    \label{fig:topology_meanflows}
\end{center}
\end{figure}

Each MHD simulation is initialized with a large-scale poloidal field described by Eqs.~\ref{eq:poloidal_field} and ~\ref{eq:poloidal_potential}. As each case evolves, the $\Omega$-effect winds up the poloidal field generating a toroidal component, whose morphology and amplitude depend on the initial topology. Figure~\ref{fig:topology_meanflows} compares three characteristic cases, one for each topology at the strongest initial field amplitude: CMHD20 (closed), OMHD20 (open), and WMHD20 (wide-open), from top to bottom. For each case, we show the azimuthally and time-averaged differential rotation, $\overline{\Omega}/2\pi$, toroidal field, $\overline{B_\phi}$, meridional circulation represented by the latitudinal velocity, $\overline{v}$, and potential temperature perturbation, $\overline{\Theta'}$, all taken over a late-time interval of each simulation.

Notice that in the C simulations, the angular-velocity develops a strong radial shear at the top of the domain. In the O simulations the differential rotation does not depart from the purely hydrodynamic baseline (compare with ~Fig.~\ref{fig:meridional_prof}). The W cases render the contours slightly more cylindrical.

The self-consistently generated toroidal field reflects the amplitude and spatial distribution of the magnetic energy set by each topology (see contour levels in Fig.~\ref{fig:topologies}).  In the closed case, $\overline{B_\phi}$ is generated both at the base of the convection zone and in the upper convection zone, being more intense in the latter. The open topology produces the field predominantly at the base, with only a faint contribution in the upper layers. In the wide-open case, the larger initial magnetic energy density yields a more intense, larger-scale toroidal field that extends from the base of the convection zone out to the outer radii.

The meridional circulation is also modified. In the C cases, it resembles the hydrodynamic simulation near the base of the convection zone, but above this region it concentrates in a thin layer within the upper convection zone. The O case is the one that most closely resembles the HD case, with the circulation confined to a somewhat thinner region in the upper convection zone. In the W case, the circulation is weaker specially at the upper convection zone.

Finally, the potential temperature perturbation must now achieve magneto thermal wind balance. Since we restrict our analysis to the first 10 years of each simulation, during which the differential rotation changes by no more than $\sim3\%$, it does not depart much from the HD case with the exception of case C, where the baroclinicity seems to be enhanced near the top boundary.  
The open topology remains similar to the HD baseline, with only a mild difference in the upper-convection-zone gradient, whereas the wide-open case modifies the potential-temperature contours relative to HD.

In general, the figure shows that the effect of the initial poloidal field is mostly reflected in the resulting toroidal field. The C case develops a strong field within the upper convection zone, the O case keeps it largely confined to the base, and the W case results in a field spreading across the full depth of the shell. As we show in the following subsections, this difference  has relevant consequences for the characteristics of the inertial modes that develop in each case.

\subsubsection{Energy Evolution and Instability Onset}\label{sec:energy_evolution}

\begin{figure*}
    \centering

    \includegraphics[width=0.95\textwidth]{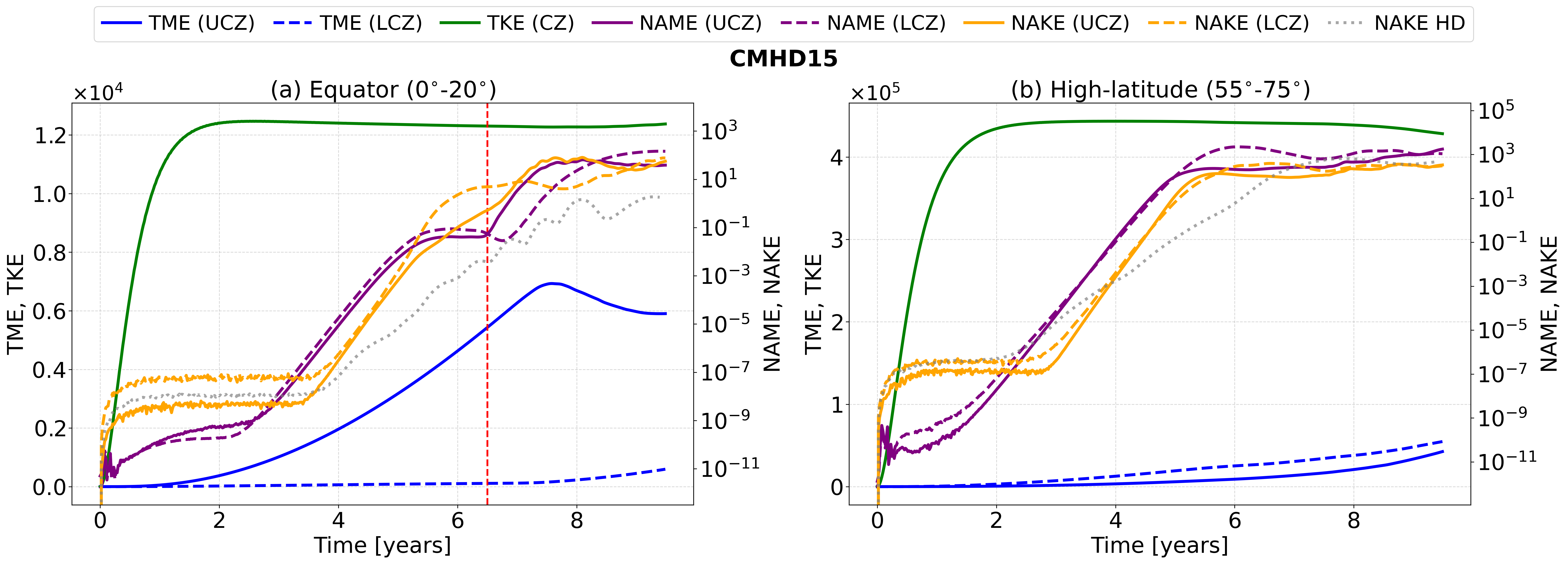}

    \vspace{0.3em}

    \includegraphics[width=0.95\textwidth]{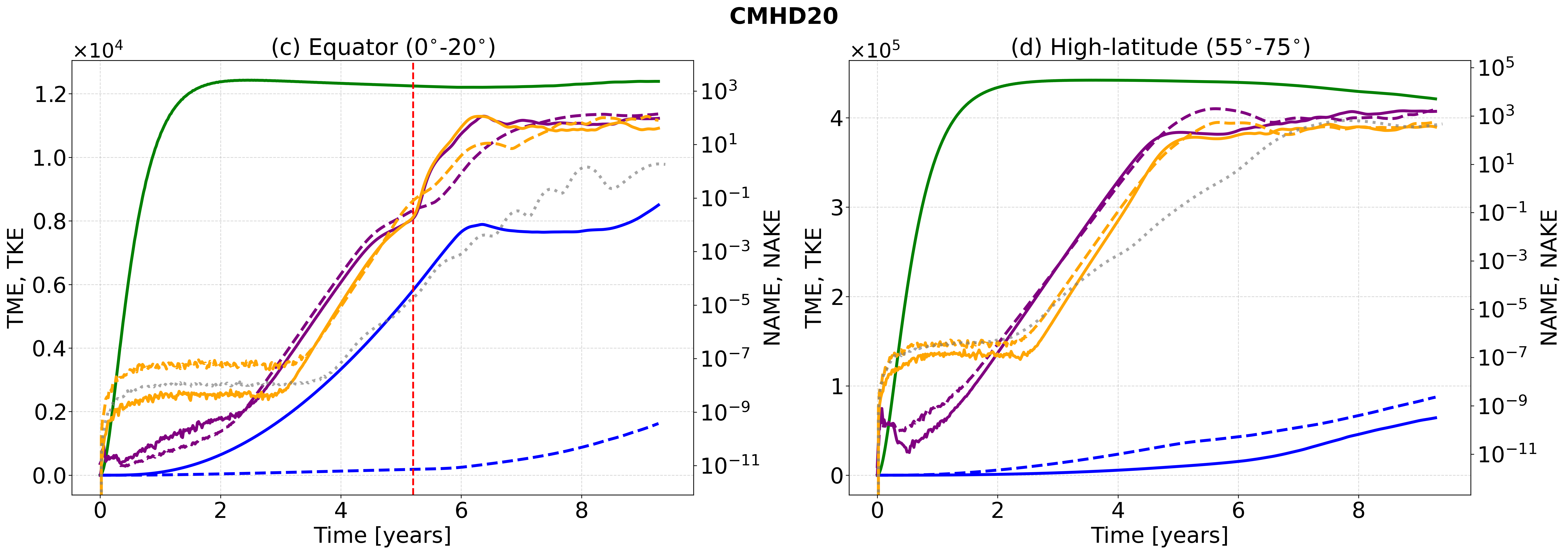}

    \vspace{0.3em}

    \includegraphics[width=0.95\textwidth]{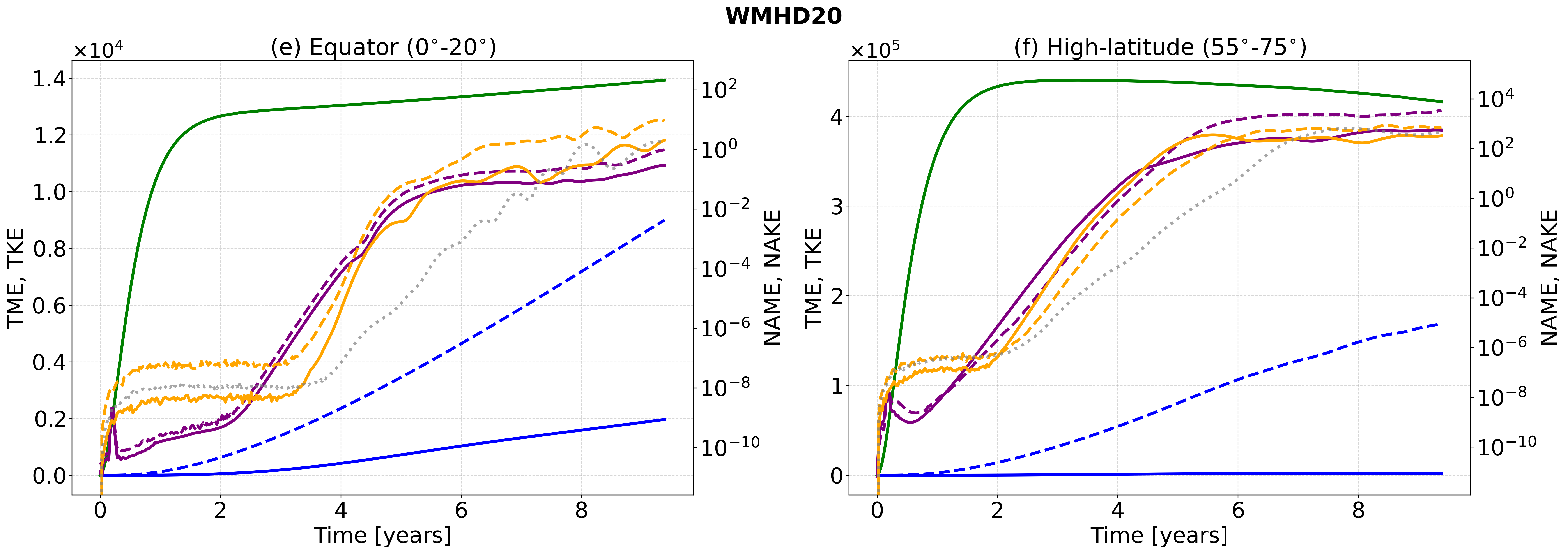}

    \caption{
    Time evolution of the energy densities for simulations CMHD15, CMHD20, and WMHD20 (from top to bottom).
    Left (right) column shows equatorial (high-latitude) region. Solid/dashed lines denote upper/lower convection-zone (UCZ/LCZ) values.
    Blue: toroidal magnetic energy density (TME); Green: toroidal kinetic energy density (TKE) averaged over the convection zone; Purple (yellow): non-axisymmetric magnetic (kinetic) energy density, NAME (NAKE); Gray dotted: hydrodynamic (HD) simulation with NAKE averaged over the convection zone. The left y-axis is shown on a linear scale and corresponds to TME and TKE,  the right y-axis is shown on a logarithmic scale and corresponds to NAME and NAKE. The vertical red dashed lines mark the onset of the second instability in panels (a) and (c).
}

    \label{fig:energy_evolution}
\end{figure*}

To characterize the energetics of the MHD simulations we decompose the flow and the field into their axisymmetric (azimuthally averaged) and non-axisymmetric ($m\neq0$) parts and define four volume-averaged energy densities. Denoting azimuthal averages by an overline  and volume averages by $\langle\rangle_V$, the toroidal kinetic energy density, $\mathrm{TKE}=\left\langle\tfrac{1}{2}\rho\overline{u}_\phi^2\right\rangle_V$, measures the energy stored in the axisymmetric longitudinal flow, i.e.\ essentially in the forced differential rotation; the toroidal magnetic energy density, $\mathrm{TME}=\left\langle\tfrac{1}{2\mu_0}\overline{B}_\phi^2\right\rangle_V$, measures the energy of the axisymmetric toroidal field developed by the $\Omega$-effect. The corresponding reservoirs of the energy of fluctuating components are the non-axisymmetric kinetic and magnetic energy densities, $\mathrm{NAKE}=\left\langle\tfrac{1}{2}\rho(u_\phi'^2+u_\theta'^2+u_r'^2)\right\rangle_V$ and $\mathrm{NAME}=\left\langle\tfrac{1}{2\mu_0}(B_\phi'^2+B_\theta'^2+B_r'^2)\right\rangle_V$, where primes denote departures from the azimuthal mean. The volume averages are computed with the spherical volume weight $r^2\sin\theta$ over selected sub-regions of the domain.

Figure~\ref{fig:energy_evolution} shows the time evolution of these energy densities for the representative cases CMHD15, CMHD20, and WMHD20 (top to bottom). Equatorial ($0^\circ$--$20^\circ$) and high-latitude ($55^\circ$--$75^\circ$) bands are presented in the left and right columns, respectively. 
As indicated by the figure legend, NAME, NAKE, and TME, represented by solid and dashed lines, distinguish the upper convection zone (UCZ, $0.91\le r/R_\odot\le0.94$) from the lower convection zone (LCZ, $0.71\le r/R_\odot\le0.74$). Conversely, TKE and the reference HD curve are averaged over the full radial extent of the convection zone ($0.71\le r/R_\odot\le0.94$). The left (linear) axis refers to TKE and TME, and the right (logarithmic) axis to NAKE and NAME; the gray dotted curve reproduces the NAKE of the HD simulation for comparison. We restrict the figure to a single wide-open case because the OMHD and WMHD cases result in nearly identical energy time evolution, so that WMHD20 is representative of both cases.

In all simulations, the TKE rises sharply within the first $\sim2$~yr as the imposed forcing establishes the differential rotation. Subsequently, the TKE evolves slowly and approximately linearly, decreasing at high latitudes in all three topologies. Its equatorial evolution, however, depends on the field topology: it increases in the O and W cases but initially decreases in the C cases. The TME initially grows monotonically as the $\Omega$-effect winds-up the poloidal field into a toroidal component. Concurrently, all simulations display an initial phase of exponential amplification of the non-axisymmetric energies, NAME and NAKE.  
This growth corresponds to the onset of the baroclinic instability already identified in the hydrodynamic regime by Paper I (see also gray dotted line). This instability sets in earlier at higher than at lower latitudes, which is consistent with the steep radial shear and the latitudinal gradient of $\overline{\Theta'}$ at tachocline levels (compare, for example, the purple lines at the left and right panels). We note that NAME begins its exponential growth slightly earlier than NAKE, yet the two display similar growth rates. This  suggests that the fluctuating field is amplified in tandem with the velocity perturbations, rather than through a distinct magnetic instability. 
The onset of the baroclinic instability depends on the field topology. For instance, in the C cases it occurs slightly earlier at the LCZ, whereas in the cases O and W it occurs slightly earlier in the UCZ. 

The most distinctive feature of the energy dynamics is a \emph{second} episode of exponential growth in the non-axisymmetric energies (see red dashed lines) which occurs only in the C cases (CMHD10, CMHD15, and CMHD20). This second growth starts at different epochs in the three C cases, at $t \simeq 9.1$, $6.5$, and $5.2$ yr for simulations CMHD10, CMHD15, and CMHD20, respectively. At these times, the toroidal magnetic field at the UCZ reaches remarkably similar configurations, i.e., latitudinal bands with roughly Gaussian distribution and maximum magnitude, $|\overline{B}_\phi|\simeq3.6$  kG. This similarity suggests that the second instability is associated with a particular magnetic field configuration of the UCZ toroidal bands.

The reduction in $d\langle|\overline{B}_\phi|\rangle/dt$ around the onset (not shown) indicates that the growing perturbations tap the toroidal magnetic reservoir in the equatorial UCZ. The equatorial TKE also declines more rapidly during this growth, suggesting that the instability draws energy primarily from the toroidal field, with an additional contribution from differential rotation. This combined response supports the magneto-shear interpretation discussed below. Although energy extraction from the toroidal field begins at the onset, the TME initially continues to increase because field generation by the $\Omega$-effect outweighs this loss. As the perturbations grow, the balance shifts: the TME reaches a local maximum and subsequently declines as energy extraction exceeds its supply by the $\Omega$-effect. After the exponential-growth phase, the equatorial TKE averaged over the convection zone begins to recover, suggesting that the resulting modes feed back on the mean flow through angular-momentum redistribution.

\begin{figure}                 
\begin{center}
    \includegraphics[width=1.0\columnwidth]{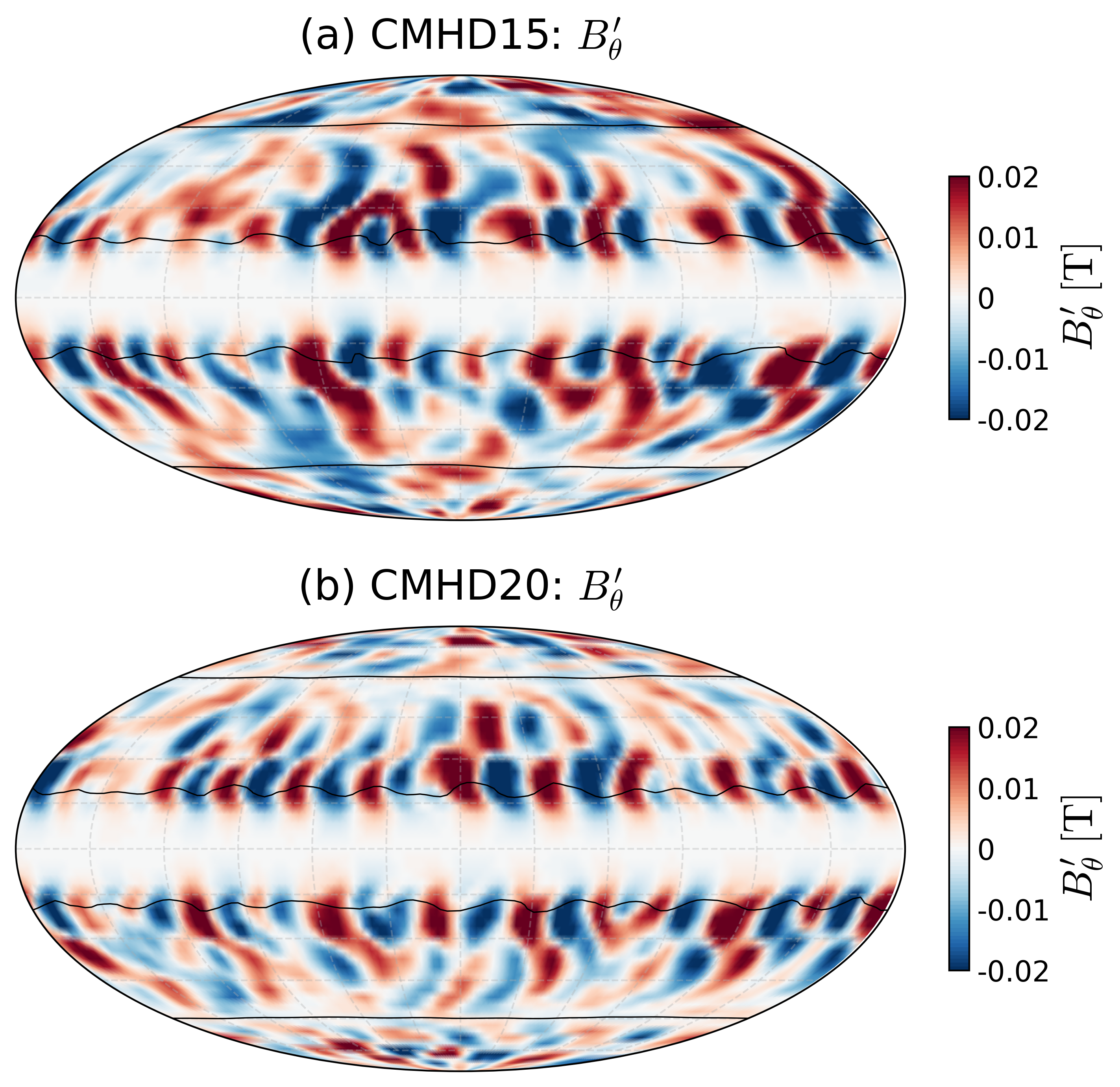}
    \caption{Mollweide projections of the non-axisymmetric latitudinal magnetic-field component, $B_{\theta}'$, at $r=0.95\,R_{\odot}$ for the CMHD15 (top) and CMHD20 (bottom) simulations. The displayed snapshots were taken at $t=7.1$ yr and $t=5.7$ yr, respectively. The color scale indicates the amplitude of $B_{\theta}'$, with red (blue) corresponding to positive (negative) values. Black contours trace $|B_\phi|$ at half the maximum magnitude of the azimuthally averaged toroidal field in each hemisphere.}
    \label{fig:mollweide_btheta}
\end{center}
\end{figure}

We associate this second instability with a magneto-shear (or ``joint'') instability of latitudinal differential rotation and toroidal field, first identified by \citet{GilmanFox1997} for broad field profiles, and by \citet{DikpatiGilman_1999, GilmanDikpati1999} for narrower banded fields. This distinction is central to our results. The instability of a broad field is restricted to azimuthal wavenumber $m=1$ \citep{GilmanFox1997}, whereas a banded field destabilizes a range of higher wavenumbers, with narrower bands favoring larger $m$ up to an optimal bandwidth \citep{DikpatiGilman_1999}. 

Figure~\ref{fig:mollweide_btheta} shows Mollweide projections of the non-axisymmetric latitudinal component of the magnetic field,  $B_\theta'$, at $r=0.95\,R_\odot$ for simulations CMHD15 at $t=7.1$~yr (panel (a)) and CMHD20 at $t=5.7$~yr (panel (b)).  Black contours trace $|B_\phi|$ at half the maximum magnitude of the azimuthally averaged toroidal field in each hemisphere, outlining the toroidal bands. The snapshots reveal a clear high-azimuthal-wavenumber pattern concentrated near the equatorward half-maximum contour during the exponential-growth phase. This contour exhibits longitudinal undulations, whereas its poleward counterpart remains comparatively smooth, particularly in CMHD20. The high-$m$ pattern in $B_\theta'$ emerges at the onset of the second instability, strengthens during exponential growth, and subsequently fades. A closely analogous signature was found by \citet{Miesch2007} in three-dimensional thin-shell simulations of the tachocline. They referred to this behavior as an ''edge instability'' of banded magnetic fields, which develops at the boundaries of the band and is most pronounced along its equatorward edge (see their Fig.~17). Thus, we believe that our findings correspond to the ``edge'' instability of toroidal field bands in the UCZ. 

We also find Alfv\'enic signatures in the closed runs. These become more prominent shortly after the second instability sets in and are clearly visible throughout the UCZ at intermediate latitudes, where the toroidal field band is most concentrated and intense.

In summary, the temporal evolution of energy for the C cases reveals the development of two different instabilities, the baroclinic instability, common to all other cases, which excites the high-latitude modes, and a magneto-shear instability triggered when the equatorward edge of the strong toroidal band in the UCZ reaches a characteristic state. The latter instability gets energy from the axisymmetric toroidal field and drives non-axisymmetric plasma motions at equatorial latitudes.  This, in turn, gives rise to the equatorial Rossby modes otherwise absent in the purely HD case. After the saturation of the non-axisymmetric modes, i.e., during the non-linear stage,  understanding the energy budget becomes much more difficult. Note that the developed helical motions and fields may result in a magnetic $\alpha$-effect in addition to the $\Omega$-effect driven by the external forcing.

\begin{figure}                 
\begin{center}
    \includegraphics[width=1.0\columnwidth]{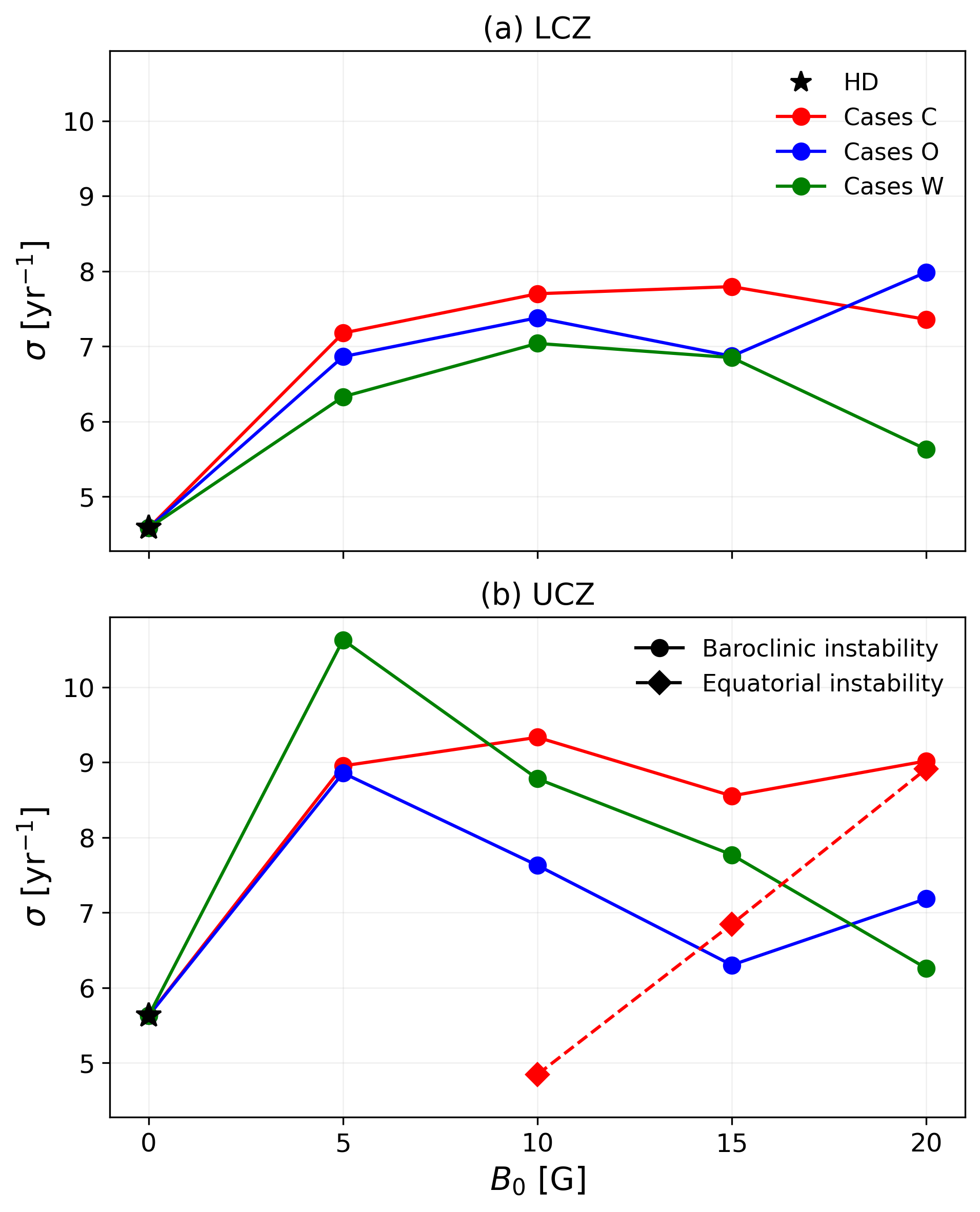}
    \caption{Growth rate, $\sigma$, of the baroclinic (solid lines) and edge (dashed) instabilities as a function of the initial poloidal field strength, $B_0$, computed in the LCZ (panel (a)) and UCZ (panel (b)). The curves compare three magnetic configurations: CMHD (red), OMHD (blue), and WMHD (green). The black star corresponds to the growth rate of the baroclinic instability in the HD case.}
    \label{fig:growthrates}
\end{center}
\end{figure}

Figure~\ref{fig:growthrates} collects the growth rates, $\sigma$, of the two instabilities described above as a function of the initial field amplitude, $B_0$, for the three magnetic topologies, C (red lines), O (blue), and W (green). Here, $\sigma$ is defined as the temporal slope of $\ln(\mathrm{NAKE})$ during the corresponding  linear-growth phase. The baroclinic growth rates are computed separately in the LCZ (panel (a)) and UCZ (panel (b)).  The second, equatorial,   instability is evaluated at lower latitudes ($0^\circ$--$20^\circ$) at the UCZ, therefore, their corresponding $\sigma$ are presented with diamonds in panel (b). This instability is found only in simulations CMHD10, CMHD15, and CMHD20,  its growth rate increases approximately linearly with $B_0$.  At the same time, stronger initial fields produce a faster buildup of the local toroidal field through the $\Omega$-effect and therefore an earlier onset of the instability. Thus, both the onset time and the subsequent growth rate of the equatorial instability show a clear dependence on the magnetic-field amplitude.

The growth rates presented in panel (a) for the LCZ and panel (b) for the UCZ reveal that the baroclinic instability depends on radial depth. The HD case provides reference growth rates of $\sigma_{\rm HD}\simeq4.6\,{\rm yr}^{-1}$ in the LCZ and $\simeq5.6\,{\rm yr}^{-1}$ in the UCZ. In both regions, introducing even the weakest magnetic field increases $\sigma$ substantially above the HD value. For stronger fields, however, the behavior depends on the magnetic topology. The C cases maintain relatively large growth rates with only moderate variations with $B_0$. The O cases exhibit different trends in the two radial regions. In the UCZ, the growth rate generally decreases from OMHD5 to OMHD15, followed by a partial recovery in OMHD20. In the LCZ, the dependence is non-monotonic, with the largest growth rate occurring for OMHD20. The W cases exhibit the clearest weakening with increasing initial magnetic-field strength in the UCZ. In the LCZ, the growth rate initially increases from WMHD5 to WMHD10 and subsequently decreases, approaching the HD value for WMHD20.

The weakening of the baroclinic instability at stronger fields, most clearly seen in the W cases, is qualitatively consistent with the linear analysis of \citet{Gilman_2015}, who found that increasing the toroidal-field strength progressively stabilizes baroclinic modes, with growth rates decreasing more rapidly at stronger fields. In our nonlinear simulations, $B_0$ specifies the initial poloidal-field amplitude, while different toroidal-field distributions develop self-consistently depending on the topology and depth. Therefore, a monotonic dependence of $\sigma$ on $B_0$ is not expected. Nevertheless, the W cases show a clear transition from an initial enhancement of baroclinic growth to a weakening at stronger fields, providing a fully nonlinear three-dimensional counterpart to the stabilizing tendency identified by \citet{Gilman_2015}, while showing that this behavior also depends on radial depth.

\subsubsection{Excited Inertial Modes}\label{sec:inertial_modes}

The energy diagnostics discussed above show that all simulations develop the baroclinic instability associated with the high-latitude modes, whereas only the C cases develop a second instability which we associate with equatorial Rossby waves. In this section we explore how these instabilities manifest in spectral space and the role of the magnetic field in the resulting inertial modes.

\begin{figure}[t]
    \centering

    \includegraphics[width=0.49\columnwidth]
        {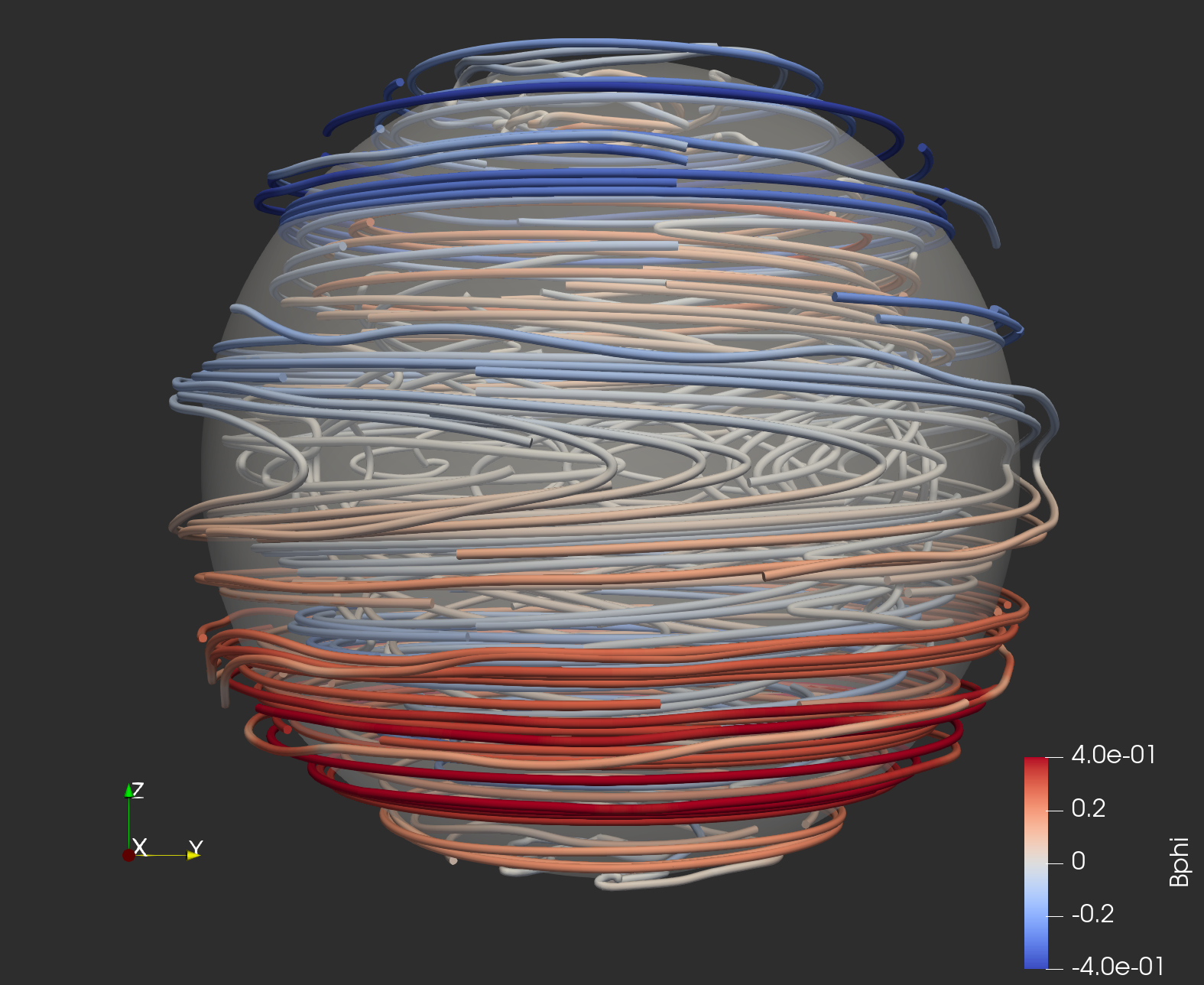}
    \hfill
    \includegraphics[width=0.49\columnwidth]
        {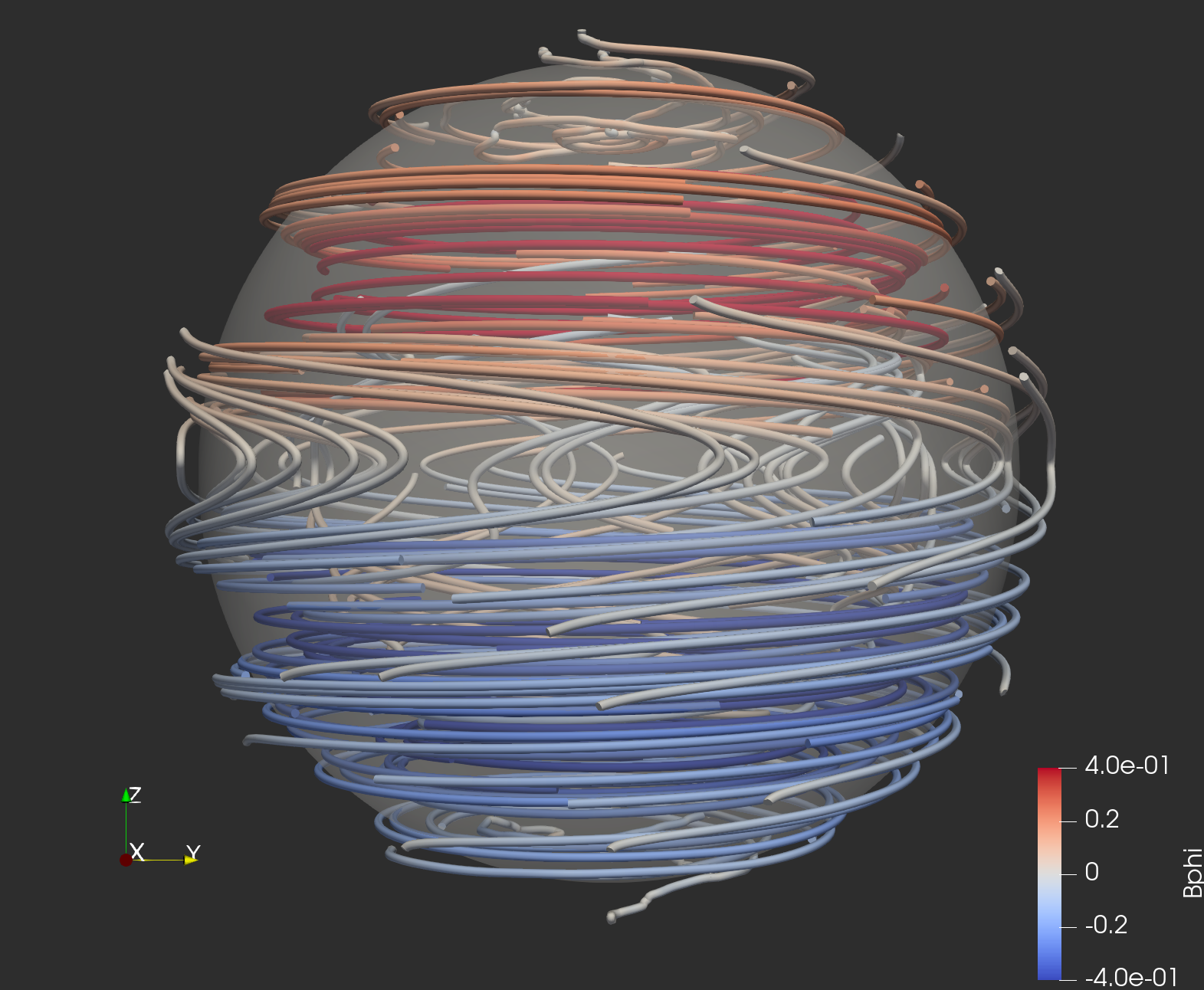}

    \par\smallskip

    \includegraphics[width=0.49\columnwidth]
        {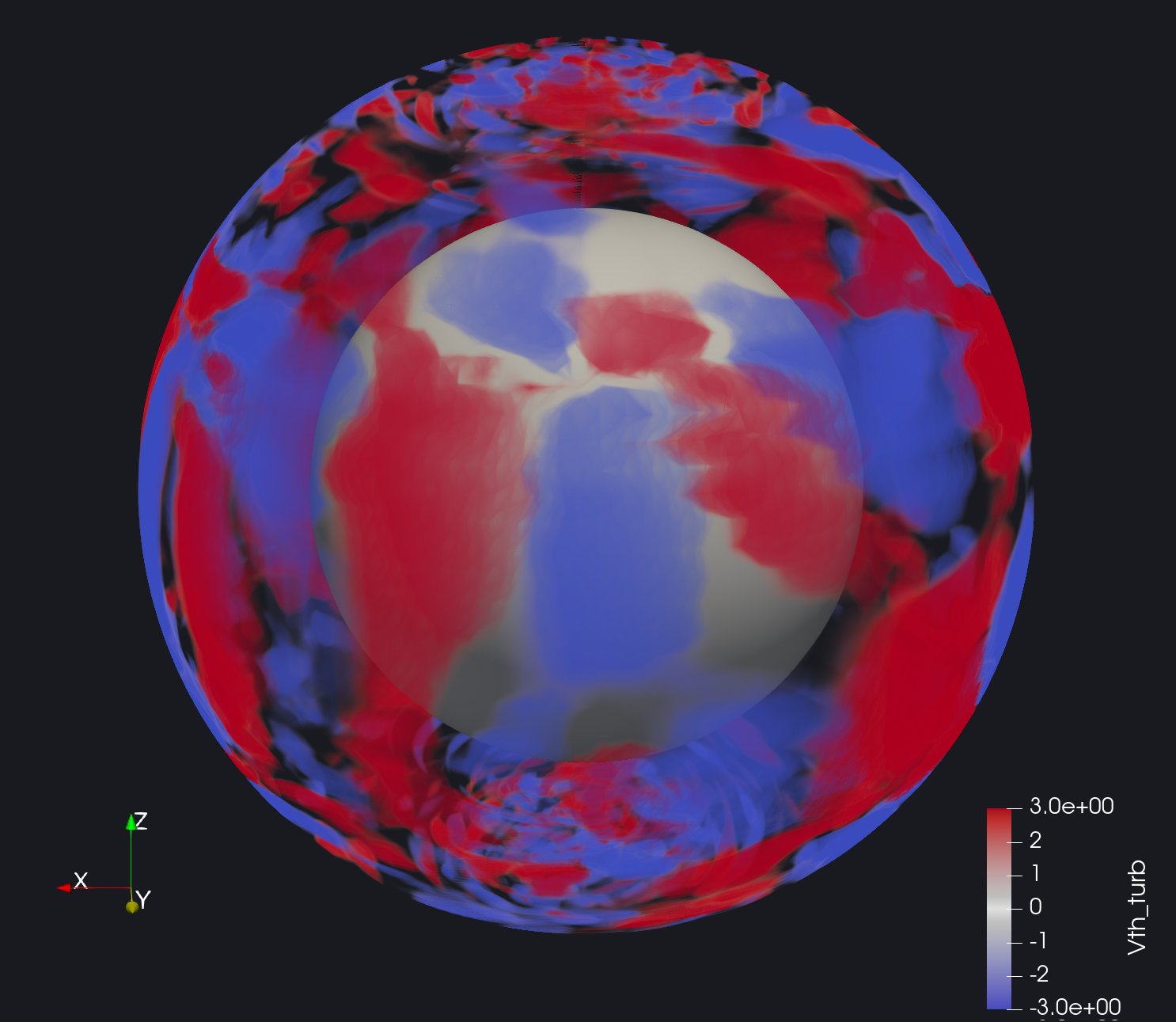}
    \hfill
    \includegraphics[width=0.49\columnwidth]
        {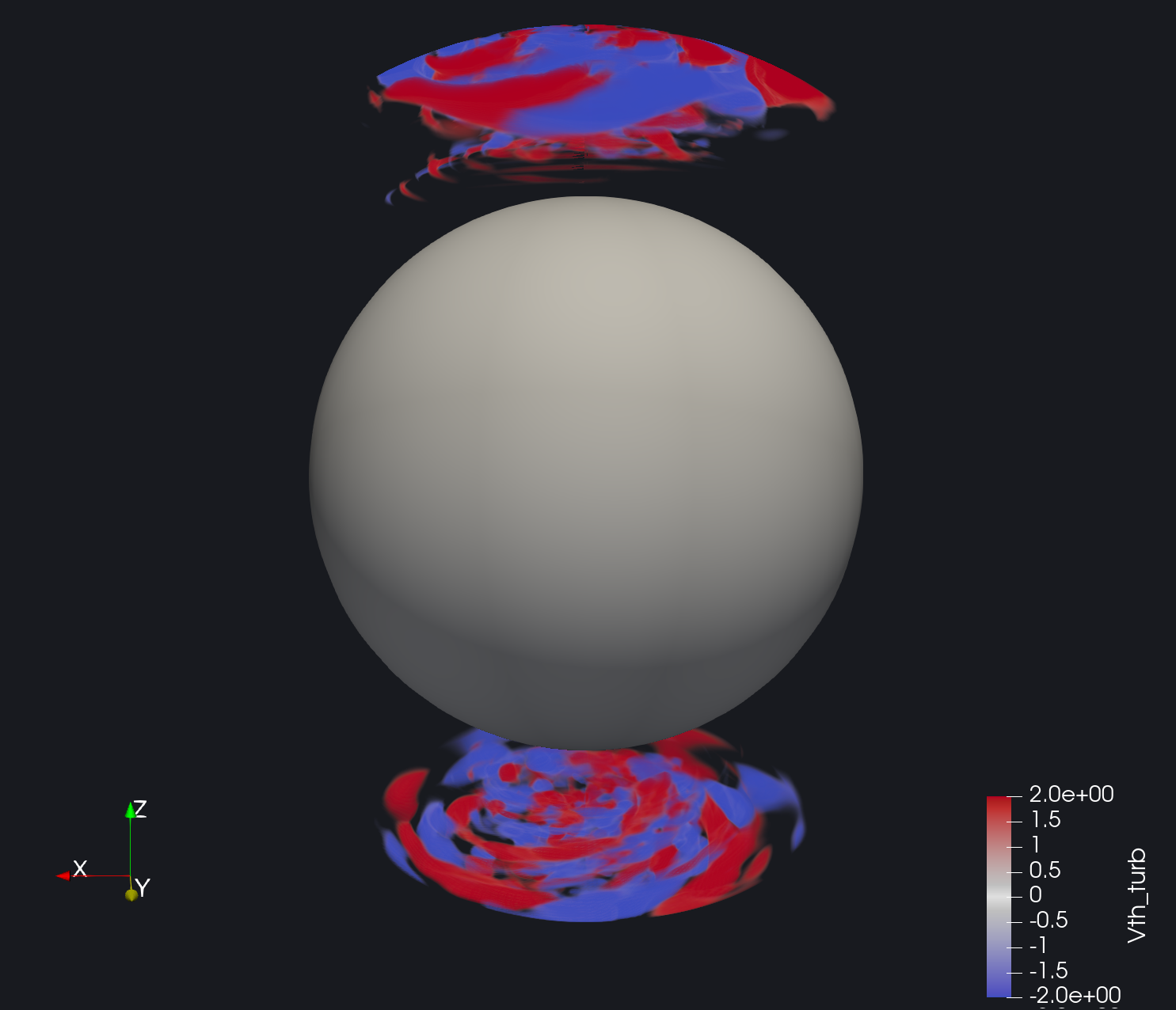}

    \caption{Snapshots of magnetic field lines (top row) and the non-axisymmetric latitudinal velocity, $v'$  (bottom), for simulations CMHD20 (left column)  and WMHD20 (right) at $t = 6$ yr.}

    \label{fig:anim_snapshots}
\end{figure}

Figure~\ref{fig:anim_snapshots} shows snapshots of the non-axisymmetric latitudinal velocity, $v'$, and magnetic field lines in CMHD20 and WMHD20. In addition to the high-latitude disturbances present in both simulations, CMHD20 develops a distinct equatorial pattern associated with the second instability. The undulating magnetic field lines in CMHD20 also reveal the signature of Alfvénic oscillations. The corresponding animations are available in the accompanying Zenodo archive \citep{Finotti2026anims}.

\begin{figure*}
\centering

\begin{subfigure}{0.495\textwidth}
\includegraphics[width=\linewidth]{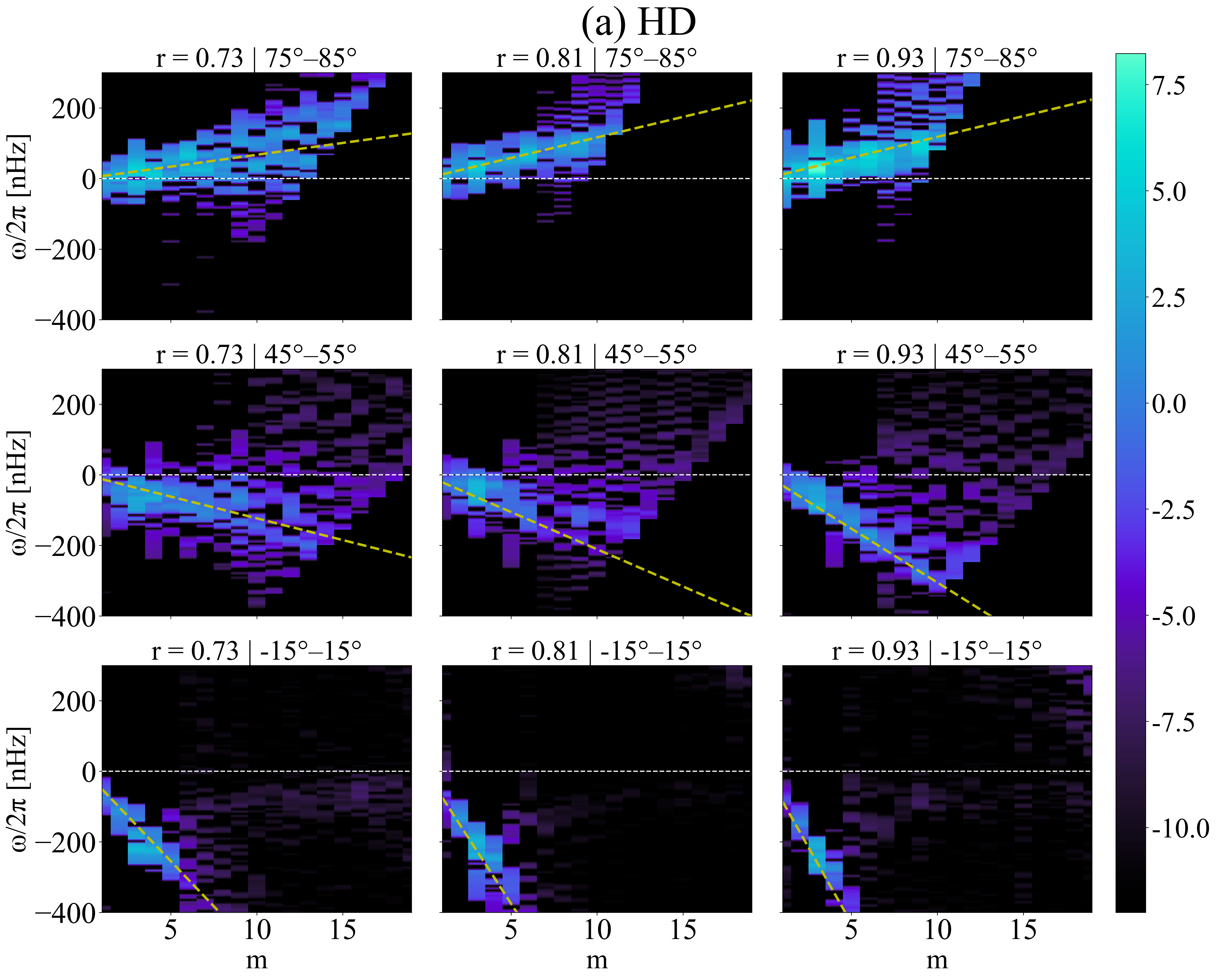}
\end{subfigure}%
\hspace{0.005\textwidth}%
\begin{subfigure}{0.495\textwidth}
\includegraphics[width=\linewidth]{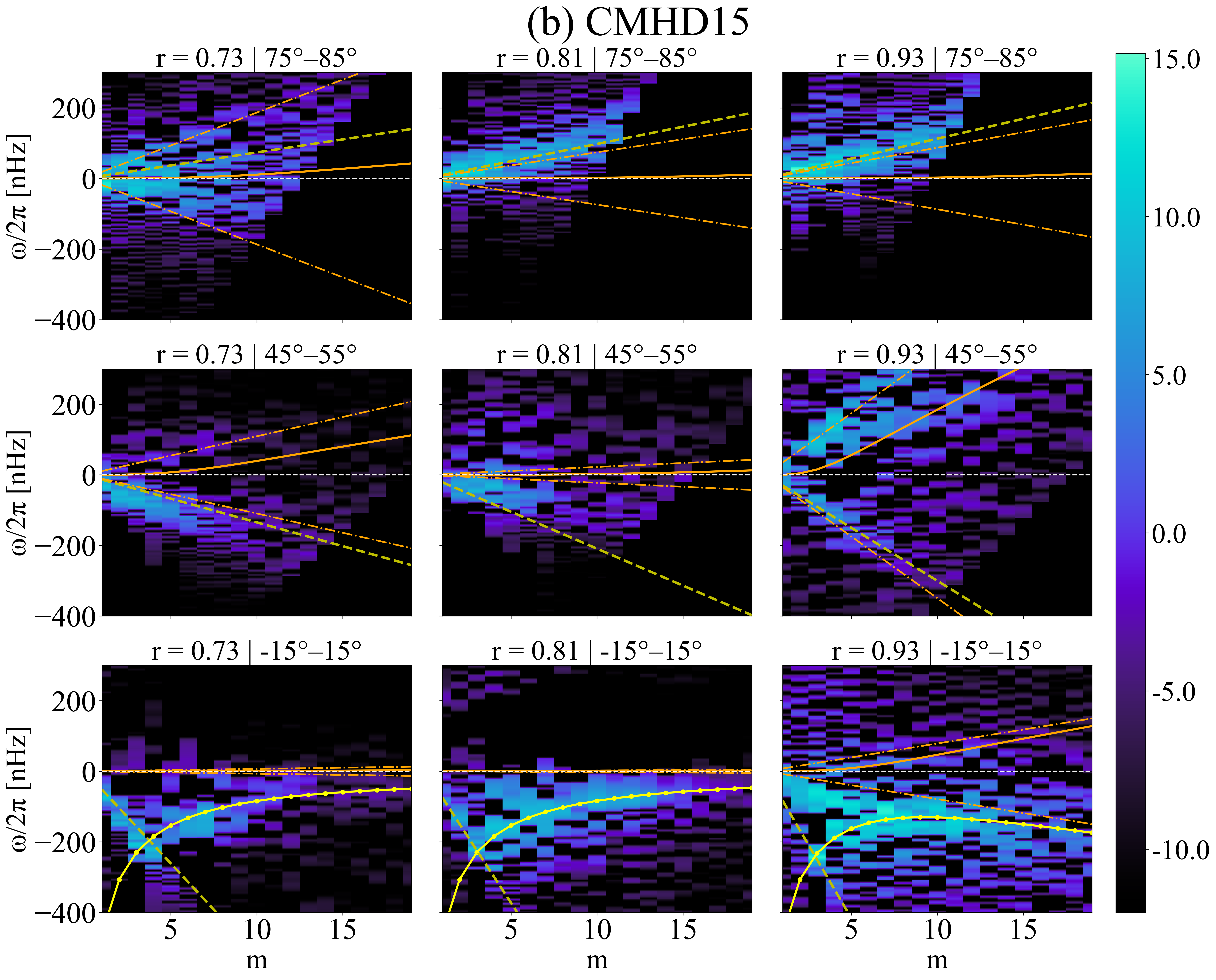}
\end{subfigure}

\vspace{0.25em}

\begin{subfigure}{0.495\textwidth}
\includegraphics[width=\linewidth]{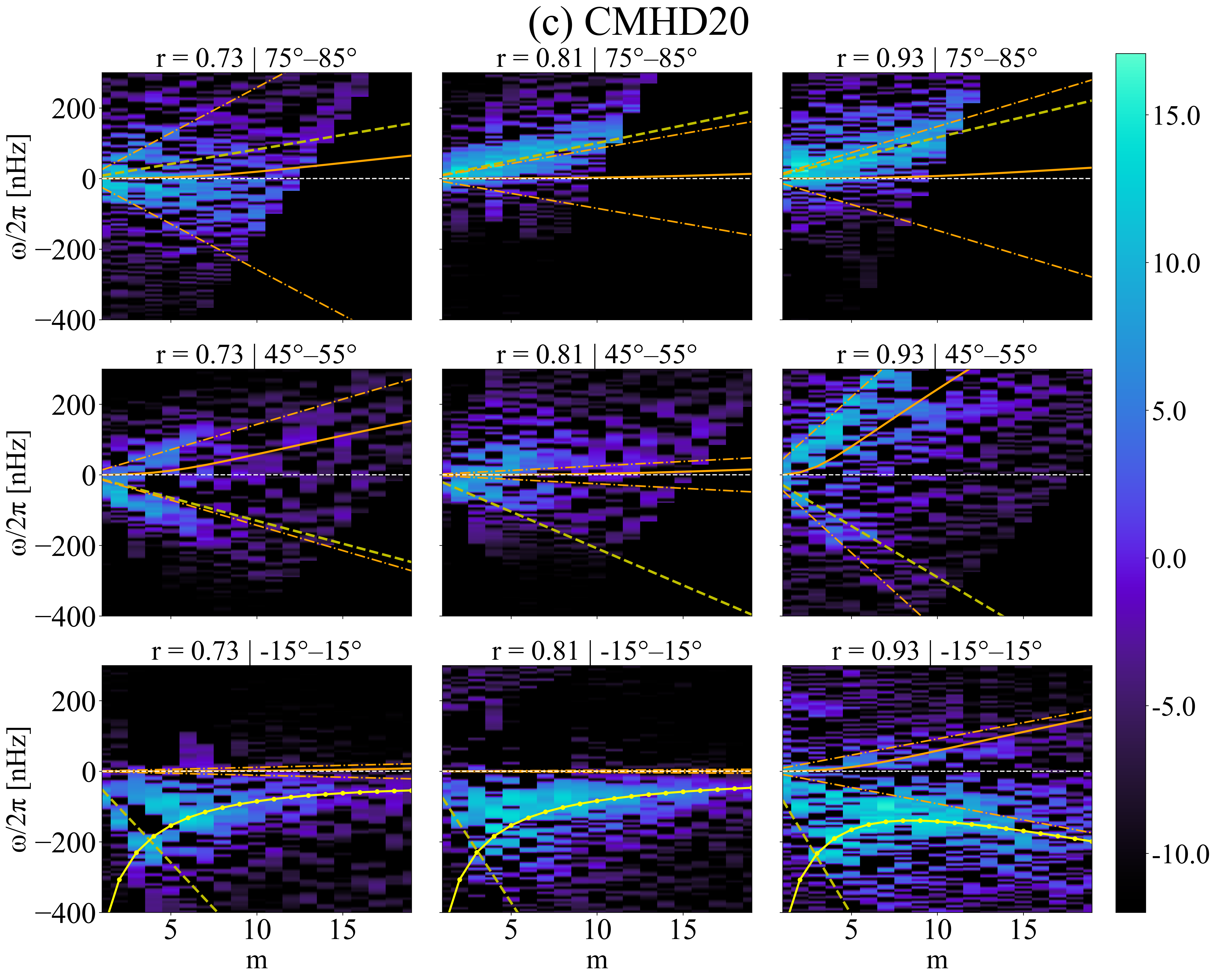}
\end{subfigure}%
\hspace{0.005\textwidth}%
\begin{subfigure}{0.495\textwidth}
\includegraphics[width=\linewidth]{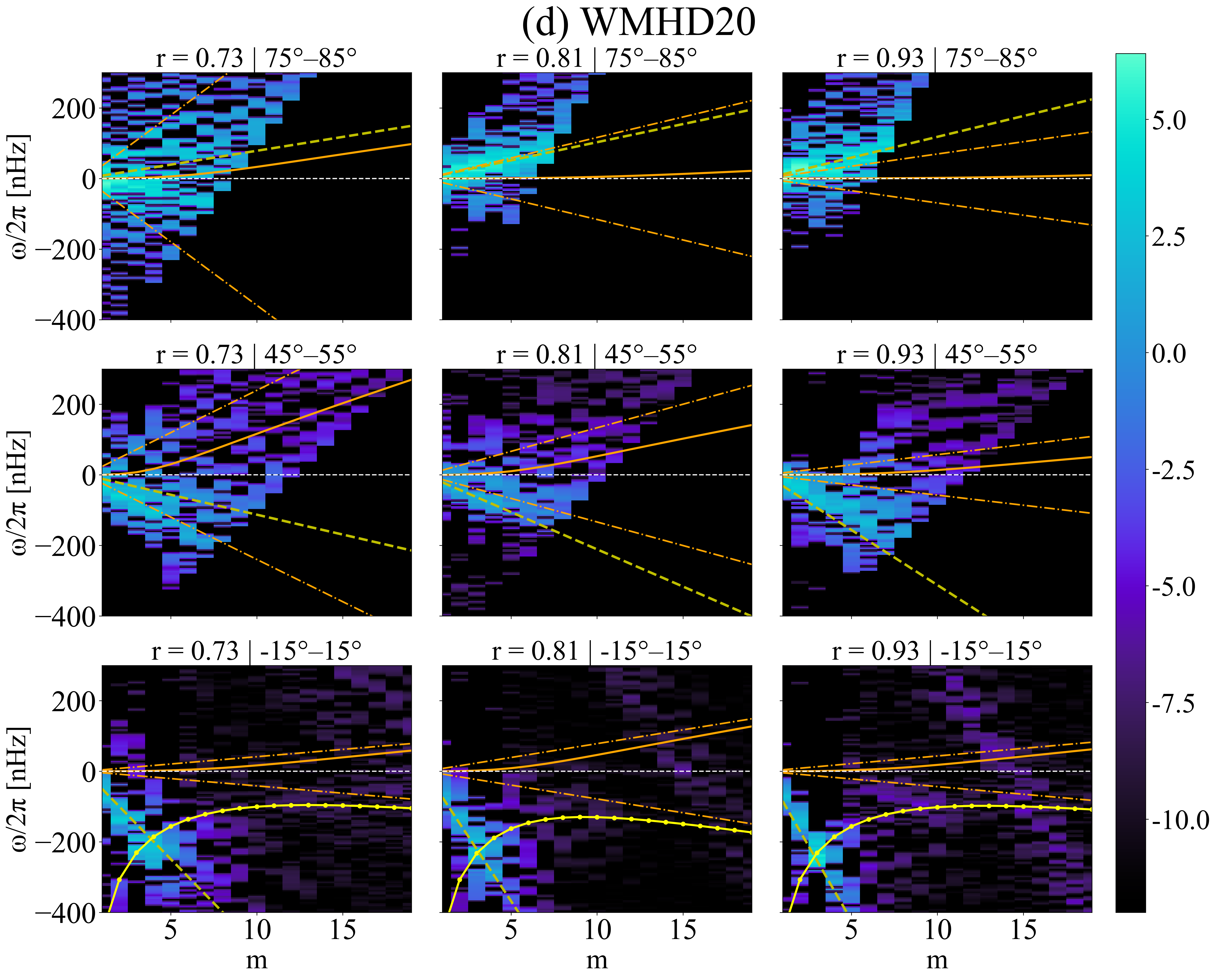}
\end{subfigure}

\caption{
Power spectra of the non-axisymmetric latitudinal velocity, $v^{\prime}$ for the simulations HD (panel a), CMHD15 (b), CMHD20 (c), and WMHD20 (d) for the ranges $-400 \,\text{nHz} \leq \omega/2\pi \leq 300 \,\text{nHz}$, and $1 \leq m < 20$.
Rows show latitudinal bands and columns show radii, as indicated in each panel. 
Yellow dashed lines show the dispersion relation of the high-latitude branch.  In the equatorial bands (third row of each panel), the yellow solid curves mark the fast magneto-Rossby branch. Orange solid lines mark the slow magneto-Rossby branch. Orange dash-dotted lines represent Alfvénic wave frequencies. The white horizontal dashed line marks $\omega=0$. 
}
\label{fig:spectra}
\end{figure*}

We construct power spectra in the  $(m,\omega)$ plane (see  Figure~\ref{fig:spectra}).  The spectra are built as follows,  the non-axisymmetric latitudinal velocity $v'$ is Fourier transformed in longitude, integrated over selected latitudinal bands, and then Fourier transformed in time. The spectra are Doppler-corrected to the local corotating frame using the mean rotation rate of each latitudinal band. To enhance coherent ridges, we subtract a running-median background in frequency and display the power as $\log_{10}(P^\Gamma)$. Since the exponent is a monotonic rescaling, it changes only the contrast, not the position or relative ordering of the spectral features. Larger values are adopted for the C cases, whose spectra are noisier: $\Gamma = 1.5$ for HD and WMHD20, $\Gamma = 3.0$ for CMHD15, and $\Gamma = 3.5$ for CMHD20. This correction affects only the visual contrast, so color intensities should not be compared quantitatively between simulations. We analyze three different depths, $r/R_\odot=0.73,0.81,$ and $0.93$, and three latitudinal bands, namely, $75^\circ\leq|\lambda|\leq85^\circ$, $45^\circ\leq|\lambda|\leq55^\circ$, and $|\lambda|\leq15^\circ$.

The overplotted curves are used as guides to identify the dominant spectral branches. For the high-latitude modes (yellow dashed lines), we plot $\omega_{\rm HL} \simeq m\Delta\Omega_{\rm p}$, where $\Delta\Omega_{\rm p}$ is the difference between the mean rotation rates of the polar band ($55^\circ < \lambda < 75^\circ$) and the latitudinal band used to compute each spectrum (see Paper I for more details).

For the equatorial modes, we use the dispersion relation of magneto-Rossby waves derived in the shallow-water MHD framework for a rotating spherical layer permeated by a toroidal magnetic field \citep{Zaqarashvili2007} as an approximate guide. In its dimensionless form, the relation can be written as
$$
l(l+1)x^2+x+\alpha^2\left[2-l(l+1)\right]=0,
$$
where $x=\omega/(2m\Omega_0)$, $\alpha=v_{\rm A}/(2\Omega_0 r)$, and $v_{\rm A}=B_{\phi,{\rm loc}}/\sqrt{\mu_0\rho}$. Here, $m$ is the azimuthal wavenumber, $l$ is the spherical-harmonic degree. The quantities $\Omega_0$ and $B_{\phi,{\rm loc}}$ denote the angular velocity and toroidal-field strength, respectively, evaluated from the simulations at the corresponding radius and within the selected latitudinal band. We adopt the sectoral approximation $l=m$, appropriate for comparison with the equatorial Rossby-like ridges. The two roots define the fast (solid yellow curves) and slow (solid orange curves) magneto-Rossby branches,

\begin{equation}
\omega_{\rm fast} = -\frac{m\Omega_0}{l(l+1)}
\left[1+\sqrt{1+4\alpha^2l(l+1)\left(l(l+1)-2\right)}\right],
\label{eq:fast_rossby}
\end{equation}
and
\begin{equation}
\omega_{\rm slow} = \frac{m\Omega_0}{l(l+1)}
\left[-1+\sqrt{1+4\alpha^2l(l+1)\left(l(l+1)-2\right)}\right].
\label{eq:slow_rossby}
\end{equation}

In the hydrodynamic limit, $\alpha\rightarrow0$, the fast branch reduces to $\omega=-2m\Omega_0/[l(l+1)]$, or $\omega=-2\Omega_0/(m+1)$ for sectoral modes, while the slow branch tends to $\omega=0$.

We also overplot the local Alfvén frequencies (orange dash-dotted lines),
$$
\omega_{\rm A}=\pm k_\phi v_{\rm A},
\qquad
k_\phi=\frac{m}{r\cos\lambda_c},
$$
where $\lambda_c$ is the central latitude of the selected band. These two solutions represent the frequencies expected for disturbances propagating azimuthally along the local toroidal field, corresponding to the prograde and retrograde Alfvén-wave branches.

The power in the HD spectra is strongly concentrated at high latitudes, where the dominant ridges closely follow the high-latitude dispersion relation. No coherent equatorial branch is visible, which is consistent with the energy diagnostics presented above. This shows that the HD simulation develops only the baroclinic instability appearing at all latitudes.

The influence of the magnetic field on the baroclinic modes is strongly localized, i.e., it is more evident at higher and intermediate latitudes.  This is evident in all MHD simulations presented in Fig.~\ref{fig:spectra}.
For the LCZ at higher latitudes, the baroclinic modes become weakened, specially for higher $m$,  when compared to the HD case. At intermediate latitudes, the baroclinic modes acquire larger retrograde frequencies than the expected by the HD dispersion relation. For the UCZ, at higher latitudes the baroclinic modes follow the same HD dispersion relation, at intermediate latitudes, they seem to be coupled with the retrograde Alfvén waves in cases CMHD15 and CMHD20 (panels (b) and (c)). The simulation WMHD20 (panel (d)) does not exhibit strong toroidal field at the UCZ.

The spectra of the MHD simulations show ridges aligned with the local Alfvén-frequency. This is specially evident in regions where the toroidal field is strongest (see the UCZ at intermediate latitudes of the C cases, panels (b) and (c)). We noticed that the prograde  Alfvén waves have larger power than the retrograde waves. Interestingly, this is the same region associated with the slow magneto-Rossby branch (see continuous orange lines). We believe that the appearance of the two kinds of waves might be related and this could be the reason for the imbalance between prograde and retrograde Alfvén waves propagation.

Unlike the HD case where no equatorial Rossby waves are observed, in the C cases (panels (b) and (c)) we observe low latitude modes associated to fast magneto-Rossby waves. The deviation from the classical HD Rossby-wave dispersion relation is most evident at  the UCZ  where the local Alfvén speed $v_{\rm A}$, and consequently $\alpha$, reaches its maximum. At the LCZ and middle of the convection zone, the power remains concentrated close to the corresponding fast-branch curves (continuous yellow lines) evaluated with the smaller local values of $v_{\rm A}$. The lower $v_{\rm A}$, the closer to the HD dispersion relation. 
The animations showing the non-axisymmetric velocity  \citep{Finotti2026anims} show that the equatorial Rossby-like pattern develops first at the UCZ and subsequently at the middle of the convection zone and LCZ.  In agreement with our discussion of the energy evolution, we conclude that a second instability originates in the strongly magnetized UCZ at equatorial latitudes, exciting magneto-Rossby modes that then propagate inwards. Nevertheless the dispersion at each depth reflects the local conditions of density and toroidal field. Note that in case WMHD20 (panel (d)), these equatorial magneto-Rossby modes do not appear.  The distribution of magnetic field in this case, does not provide conditions for the development of the second instability. This behavior is also observed in all O cases. 

In Fig.~\ref{fig:spectra} we present simulations CMHD15 and CMHD20 to highlight that the field amplitude has quantitative influence on the spectra. For instance, the magneto-Rossby and Alfvén waves are more evident in case CMHD20. Spectra of the O cases are not presented since the results are qualitatively similar to case WMHD20.

\begin{figure}[t]
    \centering
    \includegraphics[width=\columnwidth]
    {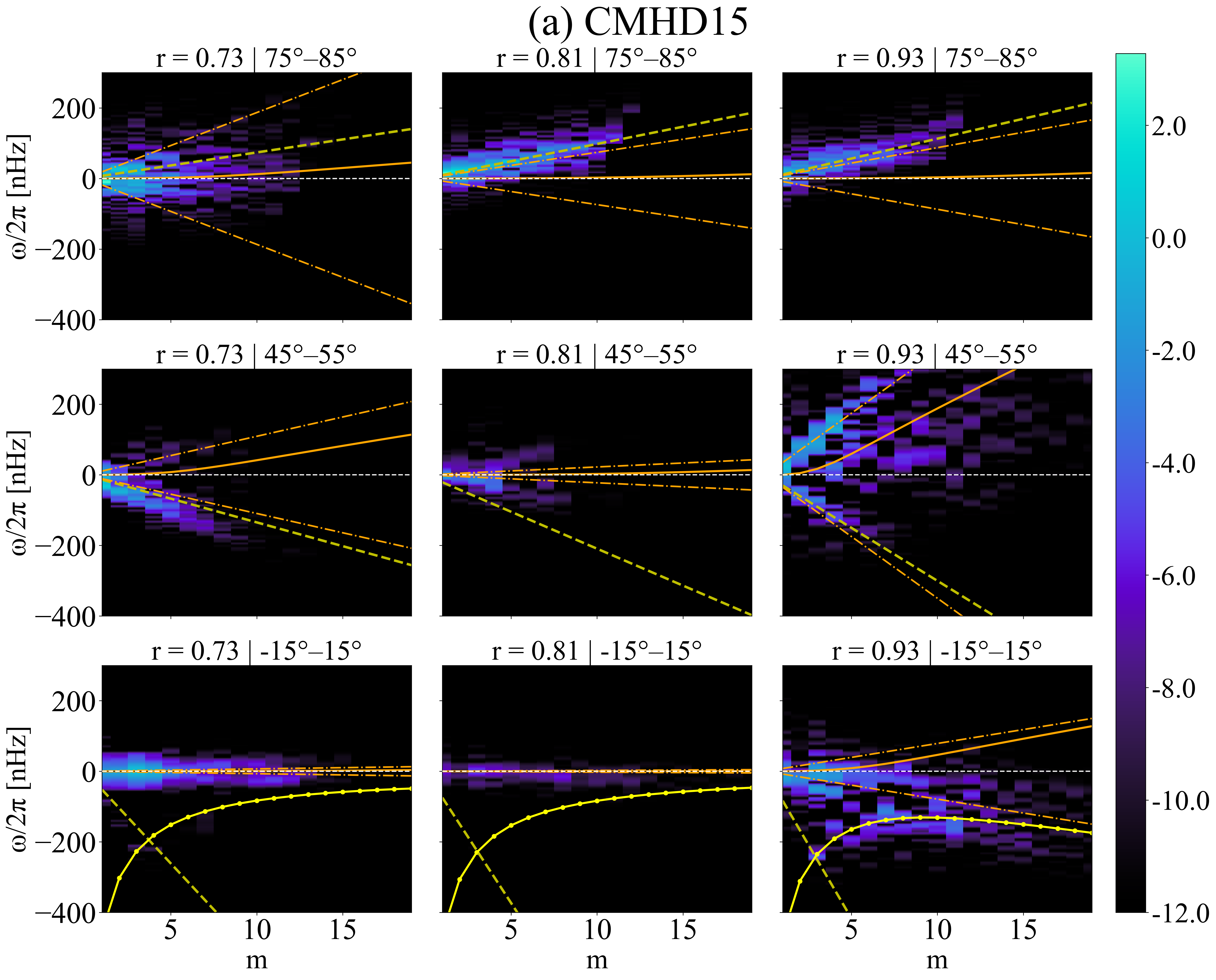}
    \par\smallskip
    \includegraphics[width=\columnwidth]
    {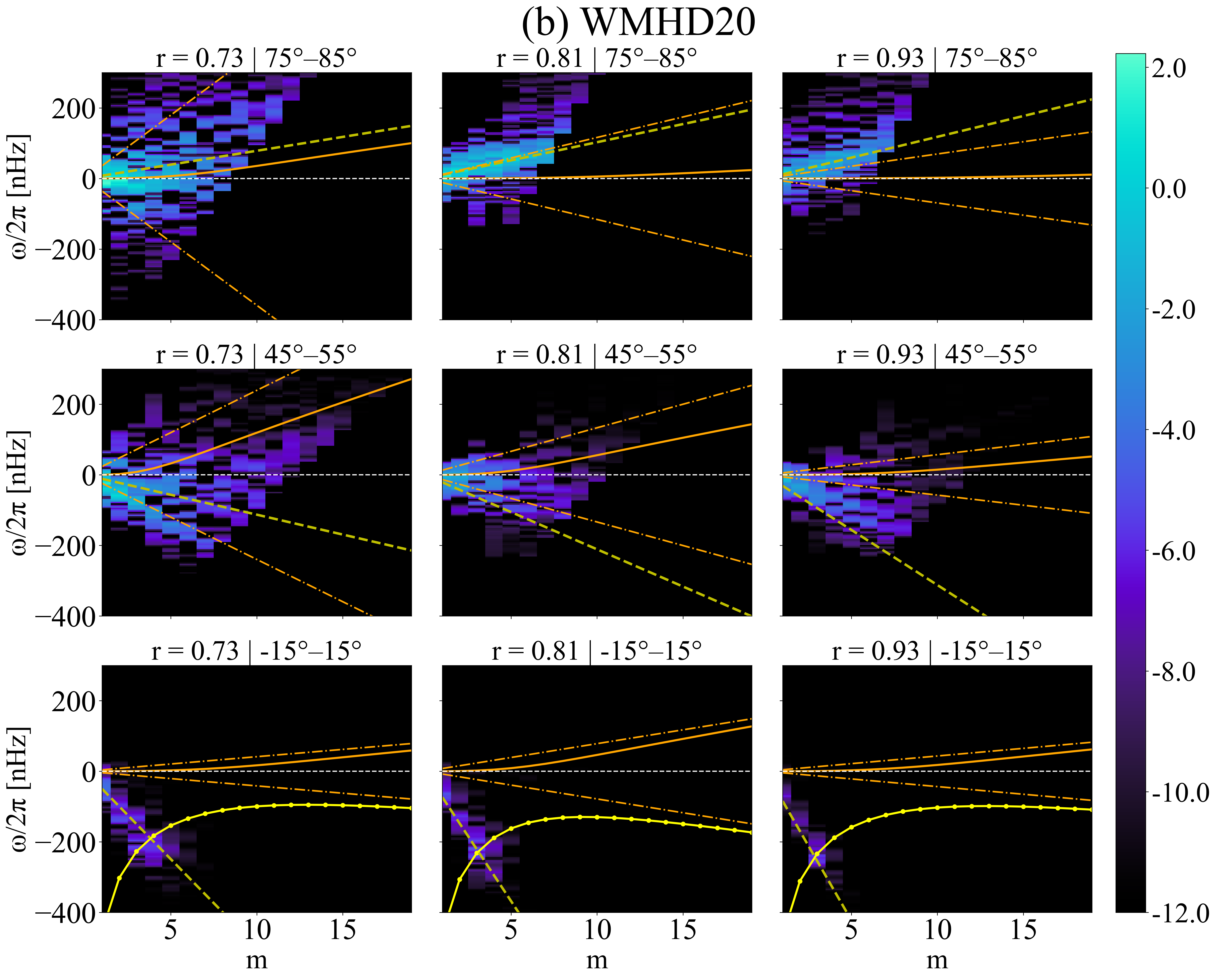}
    \caption{Same as Fig.~\ref{fig:spectra}, but in this case the spectra were computed from  the non-axisymmetric toroidal magnetic field, $B_\phi'$, in simulations CMHD15 (panel a) and WMHD20 (b).}
    \label{fig:spectra_magnetic}
\end{figure}

Figure~\ref{fig:spectra_magnetic} shows the corresponding spectra computed from the non-axisymmetric toroidal magnetic field, $B_\phi'$, for simulations CMHD15 (panel a) and WMHD20 (b). In simulation CMHD15, the equatorial UCZ exhibits a clear ridge that closely follows the fast magneto-Rossby dispersion relation. At deeper radii, however, the ridge that follows the classical Rossby dispersion relation in the velocity spectra has no coherent counterpart in the magnetic spectra. Instead, the magnetic power there is concentrated near zero frequency, consistent with the small Alfvén frequencies associated with the weaker local toroidal field. Similarly to the spectra computed for $v^{\prime}$, the imprint of the equatorial modes in the magnetic field also depends on the local conditions, specially the toroidal-field strength. In both simulations presented in the figure the high-latitude modes have identifiable signatures in the velocity and magnetic spectra, although their extensions toward lower latitudes are less evident in $B_\phi'$.

Table~\ref{tab:waves_summary} extends this comparison to the full set of simulations and summarizes the classes of waves identified in the spectra of each simulation as well as their comparison with the theoretical dispersion relations. The baroclinic waves are clearly detected in all simulations, independently of different field strengths and topologies. In contrast, Alfv\'en waves and magneto-Rossby signatures are restricted to the C cases and strongly depend on the toroidal field strength. The fast branch produces clear spectral ridges, whereas the weaker, less distinct signatures of the slow branch make its identification tentative. The classical equatorial Rossby waves  (the HD limit) are also identified in these cases in the deeper, more weakly magnetized layers where the parameter $\alpha$ (Eq.~\ref{eq:fast_rossby}) is small. Thus, the spectra show a strong dependence on local magnetic conditions. The absence of equatorial-Rossby and Alfv\'en waves in the cases O and cases W indicates that magnetic-field strength alone is insufficient to produce these modes. Their emergence appears to require a sufficiently strong toroidal field in equatorial latitudes of the UCZ. 

\begin{table*}
\centering
\caption{
Summary of the waves identified in the spectra of each simulation.
The labels ``Clear'', ``Tentative'', and ``Not detected''
describe the qualitative identification of the corresponding ridges
in the $(m,\omega)$ spectra, as well as their agreement with their
dispersion relations.
}
\label{tab:waves_summary}

\small
\setlength{\tabcolsep}{4pt}
\renewcommand{\arraystretch}{1.08}

\begin{tabular*}{\textwidth}{
@{\extracolsep{\fill}}
lccccc
@{}
}
\hline
Simulation &
\shortstack{High-Latitude \\ Baroclinic} &
\shortstack{Classical Equatorial\\Rossby} &
\shortstack{Alfv\'en} &
\shortstack{Fast Magneto-Rossby} &
\shortstack{Slow Magneto-Rossby} \\
\hline
HD     & Clear & Not detected & Not detected & Not detected & Not detected \\
CMHD5  & Clear & Not detected & Tentative    & Not detected & Not detected \\
CMHD10 & Clear & Tentative    & Clear        & Tentative    & Tentative \\
CMHD15 & Clear & Clear        & Clear        & Clear        & Tentative \\
CMHD20 & Clear & Clear        & Clear        & Clear        & Tentative \\
OMHD5  & Clear & Not detected & Not detected & Not detected & Not detected \\
OMHD10 & Clear & Not detected & Not detected & Not detected & Not detected \\
OMHD15 & Clear & Not detected & Not detected & Not detected & Not detected \\
OMHD20 & Clear & Not detected & Not detected & Not detected & Not detected \\
WMHD5  & Clear & Not detected & Not detected & Not detected & Not detected \\
WMHD10 & Clear & Not detected & Not detected & Not detected & Not detected \\
WMHD15 & Clear & Not detected & Not detected & Not detected & Not detected \\
WMHD20 & Clear & Not detected & Not detected & Not detected & Not detected \\
\hline
\end{tabular*}
\end{table*}

\subsubsection{Impact on the Solar Polar Vortex}\label{sec:polar_vortex}

In Paper I we showed that the nonlinear development of the high-latitude baroclinic instability results in a large-scale polar vortex. Although the complete non-axisymmetric longitudinal velocity displayed a clockwise spiral in the hydrodynamic simulation, filtering the flow to retain only $m=1$ and $\ell\leq22$ revealed a counterclockwise structure resembling the solar northern polar vortex. This indicated that the coherent large-scale component may be embedded within a more complex flow containing higher-order spatial scales. Here, we examine how this structure is modified by the magnetic field.

\begin{figure*}
\begin{center}
    \includegraphics[width=\textwidth]{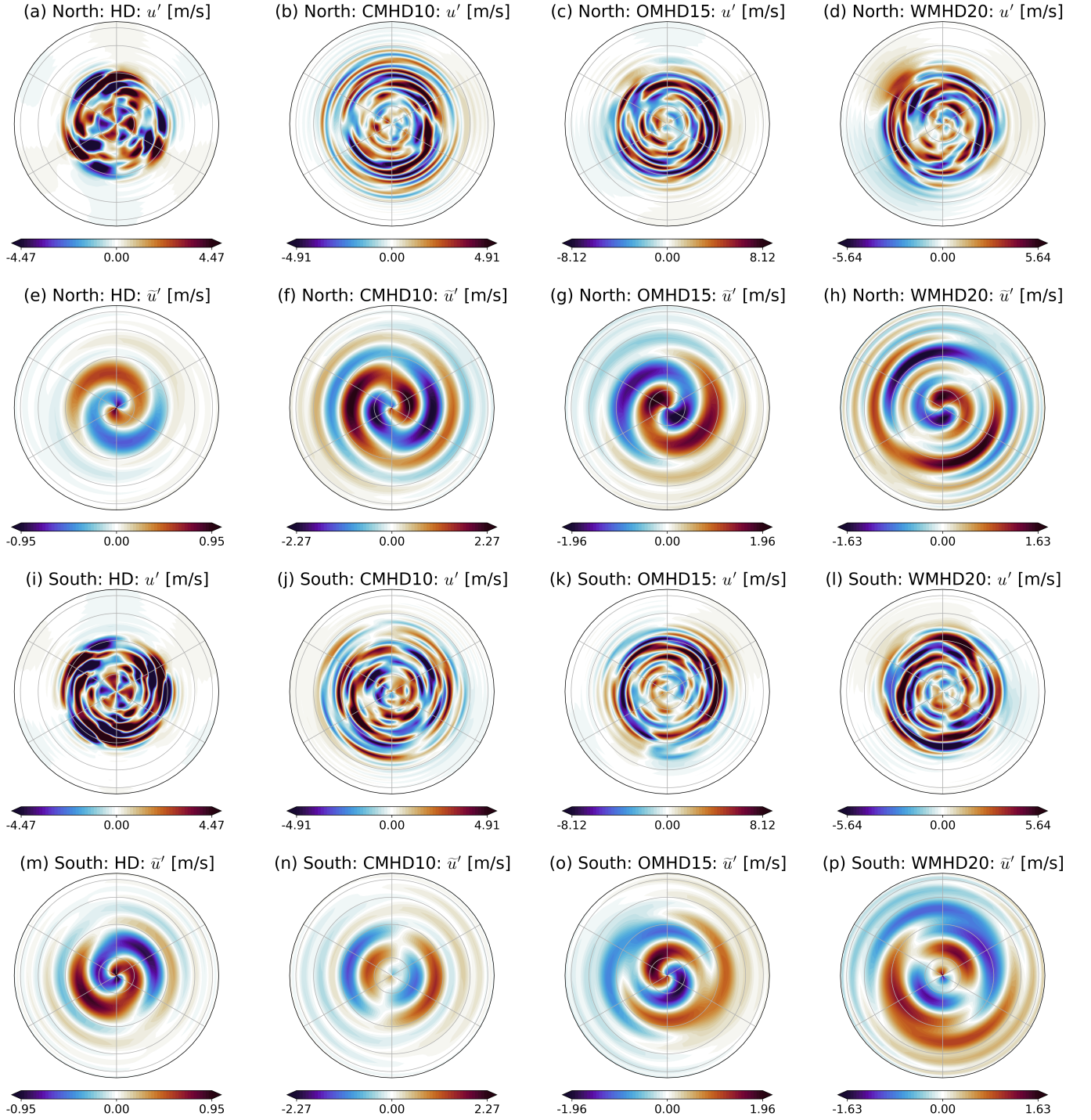}
    \caption{Snapshots of the longitudinal velocity perturbation at the north and south poles, shown in orthographic projection for simulations HD, CMHD10, OMHD15, and WMHD20 (columns from left to right), at radius $r = 0.937\,R_{\odot}$ and time $t = 8\,\mathrm{yr}$. The first and second rows show the north pole, while the third and fourth rows show the south pole. The first and third rows display the full non-axisymmetric field, $u^{\prime}$; the second and fourth rows display the filtered field, $\widetilde{u}^{\prime}$, retaining only the $m=1$ component up to $\ell_{\mathrm{max}}=22$.}
    \label{fig:polar_vortex}
\end{center}
\end{figure*}

Figure~\ref{fig:polar_vortex} shows orthographic projections of the northern and southern polar regions at $r=0.937\,R_{\odot}$ and $t=8\,{\rm yr}$ for simulations HD, CMHD10, OMHD15, and WMHD20, arranged in columns from left to right. The upper two rows show the northern polar region, and the lower two rows show the southern polar region. For each hemisphere, the full non-axisymmetric longitudinal velocity, $u'$, is displayed above its filtered counterpart, $\widetilde{u}'$, reconstructed by retaining only $m=1$ with $\ell\leq22$. Similarly to the results obtained in Paper I, the unfiltered polar
structure obtained in all simulations contains contributions from smaller scales. After filtering, the northern polar region in simulation HD depicts a counterclockwise vortex, i.e., it has the same handedness as the observed polar vortex, yet with a smaller amplitude.  Simulations CMHD10 and OMHD15 also develop  well-defined counterclockwise vortices. In particular, the one developed in simulation CMHD10 closely resembles the yin--yang morphology identified in the northern polar region by \citet{Ding2025}. Conversely, simulation WMHD20 displays a clockwise vortex. The southern filtered structures in HD and OMHD15 also exhibit well-defined spirals, with opposite handedness to their northern counterparts. In CMHD10 and WMHD20, the southern spiral orientation is less clearly defined at this time, despite the distinct northern patterns.

The animations of the filtered velocity in both polar regions \citep{Finotti2026anims} show that the spiral orientation is not always persistent. In HD, the northern vortex becomes clearly identifiable after $t\simeq6.3\,{\rm yr}$ and subsequently retains its counterclockwise handedness, while the southern vortex maintains the opposite handedness. In the MHD cases, coherent northern structures generally appear earlier, around $t\simeq4\,{\rm yr}$. In CMHD10, the northern pattern evolves toward a preferentially counterclockwise configuration. The southern vortex initially has the opposite handedness but reverses around $t\simeq8\,{\rm yr}$, acquiring the same handedness as the northern vortex. Its less distinct orientation in the snapshot is consistent with this transition. In OMHD15, both polar structures establish and retain opposite spiral handedness soon after the baroclinic instability develops. In WMHD20, both poles instead exhibit recurrent changes in orientation and complex morphology, without a persistent relation between their spiral orientations.

To estimate the relative magnitude of the large-scale structure, we define

\[
R_{u,H}
=
\frac{\mathcal{\widetilde{{U}}}_{H}}
{\mathcal{U}_{H}},
\]

where $H={\rm N,S}$, $\mathcal{U}$ represents a representative amplitude scale of $|u'|$ within the corresponding polar cap ($|\lambda|\geq50^{\circ}$), and $\widetilde{\mathcal{U}}$ denotes the corresponding filtered amplitude. 

We evaluate these diagnostics separately for each snapshot between $t=7.5$ and $8.5\,{\rm yr}$ and report their temporal means. For HD, CMHD10, OMHD15, and WMHD20, respectively, the northern ratios are $R_{u,\mathrm{N}}=0.11$, $0.30$, $0.17$, and $0.19$, while the southern ratios are $R_{u,\mathrm{S}}=0.12$, $0.19$, $0.15$, and $0.17$. The MHD cases thus exhibit larger mean filtered-to-full amplitude ratios in both polar caps, with the largest increase occurring in the northern hemisphere of CMHD10. Together with the morphological differences, these results indicate that the magnetic field modifies both the structure and the relative amplitude of the polar vortices.

Similarly, we characterize the hemispheric amplitude asymmetry as

\[
A_{\widetilde{{u}}'}
=
\frac{
\mathcal{\widetilde{U}}_{\mathrm{N}}
-
\mathcal{\widetilde{U}}_{\mathrm{S}}
}{
\mathcal{\widetilde{U}}_{\mathrm{N}}
+
\mathcal{\widetilde{U}}_{\mathrm{S}}
}.
\]
\vspace{6pt}

Whenever $A_{\widetilde{{u}}'}$ is positive (negative), the amplitude of the northern (southern) hemisphere is larger. Thus, larger (smaller) absolute values of $A_{\widetilde{{u}}'}$ indicate larger (smaller) asymmetry. The values of $A_{\widetilde{{u}}'}$ reported below are temporal means over $t=7.5$--$8.5\,{\rm yr}$. The HD simulation shows a southern preference, with $A_{\widetilde{{u}}'}=-0.19$, whereas simulation CMHD10 displays a pronounced northern asymmetry, with $A_{\widetilde{{u}}'}=0.23$. Simulations OMHD15 and WMHD20 remain much closer to hemispheric symmetry, with $A_{\widetilde{{u}}'}=0.08$ and $0.06$, respectively. In simulation CMHD10, the amplitude of the filtered velocity at the northern hemisphere is $\sim 1.6$ times its southern counterpart, which is consistent with the results reported by \citet{Ding2025}. The nearly symmetric filtered amplitudes in OMHD15 and WMHD20 therefore coexist with markedly different morphologies and temporal behavior: persistent opposite spiral handedness in OMHD15 and recurrent changes in orientation in WMHD20.

Overall, in the MHD simulations presented here, the polar vortex appears earlier than in the HD case.  This happens because the baroclinic instability develops earlier and faster in the MHD cases. The coherence, orientation, and hemispheric distribution of the vortices are also clearly dependent on the magnetic field. The toroidal field developed from a closed poloidal field, i.e., C cases, is the one that better reproduces the solar observations. However, this result is also dependent on the amplitude of the initial field. Cases with smaller $B_0$ result in more coherent vortical structures, whereas stronger fields produce more morphologically complex vortices. A similar dependence of the polar-vortex structure on magnetic-field strength was also found by \citet{Dikpati2024}.

\section{Conclusions}\label{sec:conclusions}

In this work, we have investigated the effects of large-scale magnetic fields on solar inertial modes using nonlinear, global MHD simulations. Building on the hydrodynamic study of Paper I, we adopted a differential-rotation profile constrained by helioseismic inversions and introduced large-scale poloidal magnetic fields with different strengths and topologies. Rather than prescribing a toroidal field, as is commonly done in studies of magneto-Rossby waves, the toroidal component develops self-consistently through the $\Omega$-effect. This allows the magnetic topology, differential rotation and non-axisymmetric motions to interact dynamically.

Our simulations show that the influence of magnetism on inertial modes depends strongly not only on the field strength, but also on the field topology and on the resulting radial and latitudinal distribution of the toroidal field. The closed, open, and wide-open initial poloidal configurations generate substantially different toroidal-field distributions: the closed configuration develops strong toroidal fields in the upper convection zone as well as near its base, whereas the open configuration concentrates the field predominantly near the base, and the wide-open configuration produces a more radially extended field. These differences modify the differential rotation, meridional circulation, and baroclinic structure and, in turn, shape the properties of the non-axisymmetric instabilities that develop.

An important result emerging from these simulations is that the high-latitude inertial modes remain fundamentally associated with a baroclinic instability, as found in Paper I, but their growth is significantly modified by the magnetic field. Even relatively weak magnetic fields enhance the baroclinic growth rate compared to the hydrodynamic case. As the initial field strength increases, however, the variation of the baroclinic growth rate with field strength depends on magnetic topology and depth. In particular, the wide-open configuration shows a progressive weakening of the instability at stronger fields. This non-monotonic behavior arises because $B_0$ sets only the initial poloidal-field amplitude, while the toroidal field develops self-consistently with a strength and spatial distribution that depend on the magnetic topology and radial depth. Thus, magnetic tension and baroclinicity act together in controlling the growth of the high-latitude modes rather than the magnetic field providing a simple stabilizing or destabilizing correction.

A qualitatively different behavior emerges for the closed-field configurations. Once the self-consistently generated toroidal field in the upper convection zone reaches a sufficiently strong, banded configuration, a second non-axisymmetric instability develops at low latitudes. The instability extracts energy primarily from the axisymmetric toroidal magnetic field, with an additional contribution from the differential rotation, and is concentrated near the equatorward edge of the toroidal band. Its high-$m$ structure and localization are consistent with the magneto-shear, or ``edge,'' instability previously identified for narrow toroidal-field bands. Importantly, stronger initial fields build the required toroidal configuration more rapidly and lead to both an earlier onset and a larger growth rate of this instability.

This second instability provides the dynamical pathway through which equatorial Rossby-like modes, absent in the hydrodynamic simulation, are excited in the MHD system. The closed-field cases display classical equatorial Rossby signatures together with fast and tentative slow magneto-Rossby branches and Alfv\'enic signatures, with the spectral features becoming progressively clearer as the magnetic-field amplitude increases. By contrast, these equatorial branches are not detected in the open and wide-open configurations, despite the presence of substantial magnetic energy. This demonstrates that magnetic-field strength alone is insufficient: the field must be organized so as to generate a sufficiently strong toroidal band at low latitudes in the upper convection zone. The excitation and dispersion of equatorial inertial modes, therefore, carry information about the spatial distribution and topology of the underlying magnetic field, not merely its amplitude.

Magnetism also modifies the nonlinear manifestation of the high-latitude modes in the form of the solar polar vortex. Compared with the hydrodynamic simulation, coherent large-scale polar structures generally appear earlier in the MHD cases, consistent with the earlier and faster development of the baroclinic instability. Their relative amplitude, spiral orientation, hemispheric asymmetry, and temporal persistence depend on both magnetic topology and field strength. Among the configurations considered here, the closed-field cases produce structures most similar to the observed solar polar vortex, while stronger fields tend to generate more complex and less persistent vortex morphologies.

Taken together, these results demonstrate that solar inertial modes constitute a sensitive dynamical response to the strength, topology, and location of large-scale internal magnetic fields. High-latitude modes reflect the coupled influence of magnetism and baroclinicity, whereas the appearance of equatorial Rossby, magneto-Rossby, and Alfv\'enic signatures requires a more specific magnetic configuration capable of supporting magneto-shear instability. This distinction is particularly relevant in the context of recent helioseismic evidence for magnetically modified Rossby modes. Our results suggest that observed frequency shifts cannot be interpreted in terms of magnetic-field strength alone; the field’s spatial organization and the instability responsible for exciting the mode must also be considered.

The radial dependence of the spectra provides additional insight into the origin of these magnetic signatures. \citet{HanasogeHanson2026} raised the question of whether the observed magneto-Rossby signatures retain the imprint of a magnetic field near the tachocline. In our simulations, the spectra closely reflect local magnetic conditions, approaching the hydrodynamic Rossby limit in more weakly magnetized depths. This behavior favors an interpretation in which the observed magnetic signatures primarily reflect fields within the depths sampled by the measurements. Magneto-Rossby spectra thus offer a promising diagnostic of the radial distribution of internal magnetic fields, complementing their sensitivity to field strength and topology.

More broadly, the simulations provide a bridge between simplified magneto-Rossby-wave theory and the substantially more complex dynamics of a three-dimensional, differentially rotating and stratified solar interior. Future work should explore a more realistic differential rotation, including the near-surface shear layer, as well as magnetic geometry and amplitude extrapolated from observations. Direct comparison of synthetic mode spectra and spatial eigenfunctions with helioseismic observations may then help determine the internal magnetic-field strength and topology compatible with the observed inertial-mode spectrum, advancing the prospect of using these modes as probes of otherwise inaccessible magnetic fields in the solar interior.

\begin{acknowledgments}
Computational resources supporting this work were provided by the NASA High-End Computing Program through the NASA Advanced Supercomputing Division at Ames Research Center. G.G. acknowledges financial support from CNPq grant 309695/2025-2. C.S.F. acknowledges financial support from CAPES. This work is also supported by the NSF National Center for Atmospheric Research, which is a major facility sponsored by the National Science Foundation under cooperative agreement 1852977. M.D. acknowledges partial support from various NASA grants, such as NASAHSR award 80NSSC21K1676 (awarded to NSF-NCAR), NASA-DRIVE Center award 80NSSC22M0162 to Stanford, and NASA-HSR award 80NSSC21K1678 to JHU/APL. S.T. gratefully acknowledges funding from the Belgian National Fund for Scientific Research (REFERSOC 2023/NUTATION 2024 ERC-POC FNRS project No R.E001.25/R.E001.26) and from the Belgian PRODEX program managed by the European Space Agency in collaboration with the Belgian Federal Science Policy Office (Grant No 4000150793). 
\end{acknowledgments}

\bibliographystyle{aasjournal}
\bibliography{references}

\end{document}